\documentclass{aa}  

\usepackage{graphicx}

\usepackage{txfonts}
\usepackage{amsmath,bm}
\usepackage{comment}
\begin{document}

   \title{Magnetic plasmoid explosions in the context of magnetar giant flares and fast radio bursts}

   \author{K.N. Gourgouliatos
          \inst{}}

   \institute{Laboratory of Universe Sciences, Department of Physics, University of Patras, Patras, Rio, 26504, Greece\\
              \email{kngourg@upatras.gr}
}

   \date{Received ---; accepted ---}

 
  \abstract
   {Magnetar flares are highly energetic and rare events in which intense X and $\gamma$-ray emission is released from strongly magnetised neutron stars. The events are also accompanied by mass ejection from the neutron star. Fast radio bursts are short and intense pulses of coherent radio emission. Their large dispersion measures support an extragalactic origin. While their exact origin still remains elusive, a substantial number of models associates them with strong magnetic field and high-energy relativistic plasma found in the vicinity of magnetars. There is growing evidence that some fast radio bursts are associated with flare-type events from magnetars.}
   {We provide a set of configurations describing a relativistic, spherical, magnetic plasmoid explosion. We proceed by solving the equations of relativistic magnetohydrodynamics for a system that expands while maintaining its internal equilibrium. }
   {We employed a semi-analytical approach to solve the equations of relativistic magnetohydrodynamics. We assumed self-similarity in time and radius, axial symmetry, and separation of variables in the spherical and polar angle coordinate. This allowed us to reduce the problem to solving a set of ordinary differential equations. }
   {We find the interdependent relation between pressure, mass density, Lorentz factor, and magnetic field that determines the detailed properties of the solutions. A dichotomy of solutions exists that correspond to higher and lower density and thermal pressure compared to the external one. For stronger toroidal magnetic fields, the maximum permitted expansion velocity becomes lower than the weaker toroidal fields. For a given ratio of the toroidal to the poloidal field, the inclusion of pressure and mass density leads to either a higher expansion velocity when the density and pressure are lower in regions with a higher magnetic flux or to a lower expansion velocity when the pressure and mass density are higher in regions with a higher magnetic flux.}
   {These solution classes can be applied to magnetar giant flares and fast radio bursts. Those that corresponding to overdensities and higher pressure can be associated with magnetar flares, and those corresponding to underdensities can be relevant to fast radio bursts that correspond to magnetically dominated events with low mass loading.}


   \maketitle
%

\section{Introduction}

Magnetars are hosts of highly energetic events that release radiation in X-ray and $\gamma$-rays. The most energetic of these events are giant flares, which release an isotropic luminosity up to $10^{47}$ erg/s \citep{Hurley:1999,Palmer:2005,Hurley:2005}. These events have been observed from extra-galactic sources \citep{Beniamini:2025,Trigg:2025} and have been associated with strong magnetised explosions \citep{Thompson:1995} originating from magnetars where a major reconfiguration of the global magnetic field has occurred \citep{Lyutikov:2006}. Magnetar flares also are loaded with mass entrained from the crust and the interior of the neutron star \citep{Frail:1999,Demidov:2023,Patel:2025}. Thus, a complete view of giant flares requires modelling a mass-loaded magnetised explosion, reaching relativistic velocities. Here, ‘mass-loaded’ refers to finite plasma inertia and pressure within the Magnetohydrodynamical (MHD) framework, rather than explicit source terms in the continuity equation.

The collection of an ample sample of fast radio bursts (FRBs), which are millisecond-duration transients of extragalactic origin, allows us to explore their diverse phenomenology, which indicates coherent emission from compact astrophysical sources \citep{Lorimer:2007,Petroff:2019}. Since their discovery, FRBs have challenged theoretical models, demanding mechanisms capable of releasing up to $10^{40}$ erg in radio waves over millisecond timescales, often with repetition, polarisation, and spectral properties indicative of strong magnetised environments \citep{Marcote:2020}. The growing evidence linking at least a subset of FRBs to magnetars, particularly the association of FRB 200428 with the Galactic magnetar SGR 1935+2154 \citep{Bochenek:2020, Chime:2020}, has reinforced the idea that magnetic energy release in highly magnetised neutron stars can power these events.

Despite this association, the detailed physics linking magnetic activity to FRB emission remains uncertain. The proposed mechanisms span a broad range from magnetospheric reconnection and crustal cracking to relativistic shocks in magnetar winds \citep[e.g.][]{Lyubarsky:2014,Beloborodov:2017,Metzger:2019}. Most models assume magnetically dominated outflows capable of generating coherent radio emission through synchrotron maser, shocks, or antenna processes, and possibly a combination of these \citep{Lyubarsky:2021}. Theoretical and observational developments suggest that magnetar magnetospheres may often be mass-loaded by plasma ejected during starquakes, bursts, or fallback episodes \citep{Thompson:1995, Wadiasingh:2019, Yuan:2020}. Moreover, the connection between magnetar bursts and flares and FRBs that has recently been confirmed \citep{Bochenek:2020,Tavani:2021,Beniamini:2025} stresses the necessity to consider mass-loaded magnetic detonations that might be related to an event originating from the neutron star interior and propagating to the magnetosphere \citep{Bransgrove:2025}. The presence of dense plasma in a strongly magnetised environment can substantially modify the dynamics of magnetic explosions, their ability to accelerate particles, and the conditions for coherent emission. 

We explore mass-loaded magnetic explosions as potential origins of magnetar giant flares and FRBs. Using simplified relativistic magnetohydrodynamic (MHD) models, we examine how the interplay between the magnetic field, plasma mass loading, and self-similar expansion shapes the energetics and temporal properties of such events. The proposed structures can provide an appropriate environment that can lead to the generation of short, intense events with energetics and timescales consistent with observed FRBs.

To model the coupled evolution of the magnetic field and plasma during these explosive events, we adopted the formalism developed in \cite{Gourgouliatos:2010}, which provides a self-consistent description of self-similar, co-expanding magnetised outflows. In this framework, the plasma and magnetic field evolve together under a homologous (Hubble-type) expansion, preserving the topology of the magnetic field while allowing for a time-dependent redistribution of electromagnetic and kinetic energy.  

The underlying concept traces back to the \cite{Prendergast:2005} family of self-similar, uniformly expanding, axisymmetric magnetic equilibria. These equilibria, later extended by \cite{Gourgouliatos:2008} to include non-linear poloidal currents and toroidal fields, demonstrated that magnetic structures can expand coherently, a key insight for modelling transient magnetic explosions in compact objects. Building on this foundation, \cite{Gourgouliatos:2010} formulated a relativistic configuration that couples the magnetic and velocity fields in a co-expanding MHD flow, enabling the analytical and numerical exploration of outflows that evolve self-similarly while conserving magnetic helicity and flux. In these models, the field originates from an appropriately twisted dipolar located at the origin. Due to the singularity at the origin, the field is studied beyond an inner radius that serves as a lower boundary condition. 

More recently, \cite{Barkov:2022} applied similar ideas for numerical solutions of magnetic explosions in magnetically dominated and mass-loaded environments, demonstrating that magnetic energy release can drive relativistic outflows and shocks capable of producing bright, short-duration transients. In the same work, analytical solutions of force-free magnetic fields in the absence of thermal pressure and inertia forces were presented. These solutions do not diverge at the origin. It was found that the maximum expansion rate that can be attained by the expanding structure depends on the amount of twist of the magnetic field, with highly twisted magnetic fields not being able to expand at highly relativistic rates. In the absence of any thermal pressure and inertia, the pressure on the surface of the boundary of the spherical explosion is highly anisotropic due to the magnetic field structure, that is, in the dipole case, it is maximum at the equator and becomes zero at the axis of symmetry. When a structure like this is embedded within a uniform pressure environment, it will expand sideways on the equator and will form a highly anisotropic explosion, while it will lose its internal equilibrium and coherence. The inclusion of mass density and thermal pressure leads to configurations in the form of large-scale plasmoids whose surface pressure, including the electromagnetic term, is more isotropic, and under certain conditions, it can even become completely isotropic, with the surface magnetic field completely vanishing, as was shown in the non-relativistic approach \citep{Gourgouliatos:2010b}. These structures continue their uniform expansion when they are situated in a uniform density medium and maintain their internal equilibrium \citep{Gourgouliatos:2012}.  

In the present work, we extend these ideas by embedding mass-loaded magnetic explosions within the relativistically exapanding self-similar MHD formalism of \cite{Gourgouliatos:2010}. This allows us to consistently follow the co-evolution of magnetic fields and plasma during an explosive expansion and to explore the interplay between magnetic field strength, mass, and pressure and its effect on the expansion rate. This will allow us to identify the physical conditions under which these events can generate coherent radio emission with energetics and timescales comparable to those of magnetar flares and FRBs.

This paper is organised as follows. In Sect. \ref{sec:math} we outline the physical and mathematical framework for  magnetic plasmoid explosions and describe the key ingredients of our model. Section \ref{sec:solutions}  presents the solutions and the exploration of the parameter space. The results are discussed in Sect. \ref{sec:discussion}, focusing on energy release, plasma dynamics, and physical properties of the solution. In Sect. \ref{sec:applications} we discuss the implications for astrophysical sources with a main focus on magnetar bursts and FRBs.  We conclude in Sect. \ref{sec:conclusions} and outline the limitations of the current model and the prospects for future work.

\section{Mathematical formulation}
\label{sec:math}

We considered a fluid of rest-mass density $\rho$ and pressure $p$ containing a magnetic field $\bm{ B}$. The electric and magnetic field obey Maxwell's equations,
\begin{eqnarray}
\nabla \cdot \bm{ B}&=&0\,, \label{eq:Max1} \\
\nabla \cdot \bm{ E}&=&\frac{4\pi}{c} j^0\,, \label{eq:Max2}\\
\nabla \times \bm{ E} &=&-\frac{1}{c}\frac{\partial \bm{ B}}{\partial t}\,, \label{eq:Max3}\\
\nabla \times \bm{ B} &=&\frac{1}{c}\frac{\partial \bm{ E}}{\partial t}+\frac{4\pi}{c} \bm{ j}\,, \label{eq:Max4}
\end{eqnarray}
where $j^0$ is the electric charge density, $\bm{ j}$ is the electric current density, and $c$ is the speed of light. We further considered ideal MHD, and there is thus no electric field in the co-moving frame of the fluid. For an observer who sees the fluid moving at velocity $\bm{ v}$, normalised to the speed of light, Ohm's law becomes
\begin{eqnarray}
    \bm{ E}= -\bm{ v} \times \bm{ B}\,.
    \label{eq:Ohm}
\end{eqnarray}
The continuity equation reads
\begin{eqnarray}
   \partial_t \left(\Gamma \rho \right)+\nabla \cdot \left( \Gamma \rho c\bm{v}\right)=0\,,
   \label{eq:cont}
\end{eqnarray}
where $\Gamma$ is the Lorentz factor,
\begin{eqnarray}
    \Gamma= \frac{1}{\sqrt{1-v^2}}.
\end{eqnarray}
The momentum equation is
\begin{eqnarray}
    \Gamma \rho \left(\partial_t +c\bm{v}\cdot \nabla\right)\left(\xi \Gamma c\bm{v}\right)+\nabla p -\frac{j^0 \bm{E}+\bm{j}\times\bm{B}}{c}={\bm 0}\,,
    \label{eq:mom}
\end{eqnarray}
where $p$ is the pressure, and the relativistic specific enthalpy is given by the following expression:
\begin{eqnarray}
    \xi=1+\frac{\gamma}{1-\gamma}\frac{p}{\rho c^2}\,,
\end{eqnarray}
where $\gamma$ is the ratio of the specific heats. Furthermore, the entropy equation reads
\begin{eqnarray}
    \left(\partial_t +c\bm{ v}\cdot \nabla\right)\left(\frac{p}{\rho^{\gamma}}\right)=0\,.
    \label{eq:entropy1}
\end{eqnarray}
In what follows, we assume a relativistic degenerate gas where $\gamma=4/3$. We further considered spherical coordinates $(r,\theta,\phi)$ and prescribed a velocity  normalised to the speed of light given by the following expression: 
\begin{eqnarray}
    \bm{v}= \frac{r}{ct}\hat{\bm r}\,.
\end{eqnarray}
This velocity corresponds to uniform expansion from the centre. Thus, each fluid element moves at constant velocity without any acceleration, and the system remains implicitly at force balance.  Here and throughout the paper, $v$ denotes the dimensionless radial expansion velocity $v=r/(ct)$. 

It is possible to find self-similar, separable solutions using the polar angle $\theta$ and the expansion velocity $v$ as main variables. The full details of this procedure were shown by \cite{Prendergast:2005,Gourgouliatos:2008,Gourgouliatos:2010}, and we summarise the basic steps for clarity and consistency below. 

We express the magnetic field in terms of two functions: one related to the poloidal field $\Psi(\theta,v)$, and the other to the toroidal $T(\theta,v)$,
\begin{eqnarray}
\bm{B}= \frac{1}{2 \pi r^2 \sin\theta}\left(\partial_{\theta} \Psi \hat{\bm{r}}-v\partial_{v} \Psi \hat{\bm{\theta}}+T\hat{\bm{\phi}}\right)\,.
\label{eq:mag}
\end{eqnarray}
Combining Ohm's law, Eq. (\ref{eq:Ohm}), with the momentum equation, Eq. (\ref{eq:mom}), and taking the toroidal component, we deduce that the poloidal and toroidal functions are related to each other through the following expression:
\begin{eqnarray}
    T =\Gamma^2 v H(\Psi)\,,
\end{eqnarray}
where $H(\Psi)$ is an arbitrary function of $\Psi$. 

The pressure and the density are related through the entropy equation, Eq. (\ref{eq:entropy1}), which is satisfied for
\begin{eqnarray}
    \frac{p}{\rho^{4/3}}=Q(v,\theta)\,,
    \label{eq:entropy2}
\end{eqnarray}
where $Q(v,\theta)$ is an arbitrary function of $v$ and $\theta$. Upon integration of the continuity equation, Eq. (\ref{eq:cont}), we obtain that
\begin{eqnarray}
    \rho =\tilde{\rho}(v,\theta) \frac{\Gamma^3 v^3}{r^3}\,,
    \label{eq:density}
\end{eqnarray}
and combining this with Eq. (\ref{eq:entropy2}), we find that the pressure has the following form:
\begin{eqnarray}
    p=\tilde{p}(v, \theta)\frac{\Gamma^4 v^4}{r^4}\,,
\end{eqnarray}
where the quantities $Q$ $\tilde{p}$, $\tilde{\rho}$, all depending on the separation variables $v$ and $\theta$, are related to each other through the following expression:
\begin{eqnarray}
    Q=\frac{\tilde{p}}{\tilde{\rho}^{4/3}}\,.
    \label{eq:Q}
\end{eqnarray}
Next, we evaluate the current and charge densities through Maxwell's equations, Eq. (\ref{eq:Max2}) and (\ref{eq:Max4}), and substitute them into the momentum equation, Eq. (\ref{eq:mom}). We find that $\tilde{p}=\tilde{p}(\Psi)$ and that the system should satisfy the following equation:
\begin{eqnarray}
    &v^{2} \frac{\partial^{2} \Psi}{\partial v^{2}}
- \frac{2 v^{3}}{1 - v^{2}} \frac{\partial \Psi}{\partial v} 
+ \frac{\sin\theta}{1 - v^{2}} \frac{\partial }{\partial \theta}
    \left( \frac{1}{\sin\theta} \frac{\partial \Psi}{\partial \theta} \right)\nonumber\\
&+ \frac{v^{2}}{(1 - v^{2})^{2}} H \frac{d H}{d\Psi}
+ 16 \pi^{3} \sin^{2}\theta \, \frac{v^{4}}{(1 - v^{2})^{3}} \frac{d\tilde{p}}{d\Psi}
= 0.
\label{eq:mom2}
\end{eqnarray}
Separable solutions in $v$ and $\theta$ are feasible under the following assumptions:
\begin{eqnarray}
    \Psi=g(v)\sin^2\theta\,,~H\frac{dH}{d\Psi}=c_0 \Psi\,,~ \frac{d \tilde{p}}{d\Psi}=\frac{c_1}{16\pi^3}\,.
    \label{eq:relations}
\end{eqnarray}
The function $g(v)$ determines the radial distribution of the poloidal magnetic flux under self-similar expansion. Its first non-zero root, $v_0$, defines the outer boundary of the magnetised plasmoid, beyond which all field components vanish or become disconnected from the inner region. The overall shape of $g(v)$ therefore controls both the spatial extent of the configuration and the presence or absence of surface currents and pressure anisotropies.

Upon integration, and considering the fact the $B_{\phi}$ component of the field must remain finite,
\begin{eqnarray}
H(\Psi)= \pm \sqrt{c_0} g \sin^2\theta\,,    
\end{eqnarray}
with $c_0$ a non-negative constant, which governs the intensity of the toroidal field, or equivalently, the poloidal current. The pressure term becomes
\begin{eqnarray}
    \tilde{p}= \tilde{p}_0 +\frac{c_1}{16 \pi^3} g\sin^2\theta\,.
    \label{eq:pressure}
\end{eqnarray}
In the above expression, $\tilde{p}_0$ is the pressure on the axis and the spherical surface of the plasmoid; and it has to be a non-zero constant. There is no constraint on the sign of $c_1$, but the combination of $\tilde{p}_0$ and $c_1$ must ensure that $\tilde{p}$ remains non-negative throughout the solution. 

Finally, substituting the expressions from Eq. (\ref{eq:relations}) into Eq. (\ref{eq:mom2}), we obtain an ordinary differential equation for $g(v)$,
\begin{eqnarray}
    v^{2} g^{\prime \prime} 
- \frac{2 v^{3} g^{\prime}}{1 - v^{2}}
- \frac{2 g}{1 - v^{2}}
+ \frac{c_{0} v^{2} g}{(1 - v^{2})^{2}}
+ \frac{c_{1} v^{4}}{(1 - v^{2})^{3}}
= 0\,.
\label{eq:g}
\end{eqnarray}
The first three terms describe the poloidal magnetic structure under relativistic expansion, the fourth term represents the contribution of the toroidal magnetic field (poloidal current), and the final term accounts for pressure and inertia effects.
The solution of the above equation provides the radial structure of the configuration.

\section{Solutions}

\label{sec:solutions}
The behaviour of the solution near the origin can be inspected for $v\ll 1$. The leading terms of the equation are the following:
\begin{eqnarray}
    v^2g^{\prime \prime}-2g=0\,.
\end{eqnarray}
This admits solutions in following form:
\begin{eqnarray}
g(v)= A_1v^2+A_2v^{-1}  \,,  
\end{eqnarray}
where $A_1$ and $A_2$ are constants. Thus, there are two main classes of solutions of Eq. (\ref{eq:g}). A class where $g \propto v^{-1}$, leading to solutions that reduce to a magnetic dipole near the origin, and a class where $g\propto v^2$, where the field near the centre is approximately uniform. While the former case was explored in detail in \cite{Gourgouliatos:2010}, the latter has not been studied. 
In this work, we focus on the latter class of solutions, which are not singular at the origin, by exploring appropriate combinations of the main parameters $c_0$ and $c_1$.

\subsection{Non-relativistic limit}
 In the non-relativistic limit, while keeping the terms proportional to $c_0$ and $c_1$, Eq. (\ref{eq:g}) reduces to
\begin{eqnarray}
    v^2g^{\prime \prime}-2g+c_0v^2g+c_1v^4=0\,.
\end{eqnarray}
The solution of the above equation is
\begin{eqnarray}
    g(v) &=& B_1\left(\frac{\sin\left(\sqrt{c_0} v\right)}{\sqrt{c_0 }v}-\cos\left(\sqrt{c_0}v\right)\right)\nonumber \\
    &+&B_2\left(\frac{\cos\left(\sqrt{c_0} v\right)}{\sqrt{c_0 }v}+\sin\left(\sqrt{c_0}v\right)\right)-\frac{c_1}{c_0}v^2\,,
\end{eqnarray}
where $B_1$ and $B_2$ are constants. The term multiplied with $B_1$ is zero at the origin, whereas the term multiplied with $B_2$ is singular. Therefore, we kept the $B_1$ terms and set $B_2=0$. This solution is mathematically identical to the solutions found in \citet[Eq. 6]{Gourgouliatos:2010b}, where the role of $v$ is taken by $r$. In the present case, the sphere instead of being static, it expands uniformly. 

\subsection{Relativistic force-free limit}

\begin{figure}
     \includegraphics[width=0.45\textwidth]{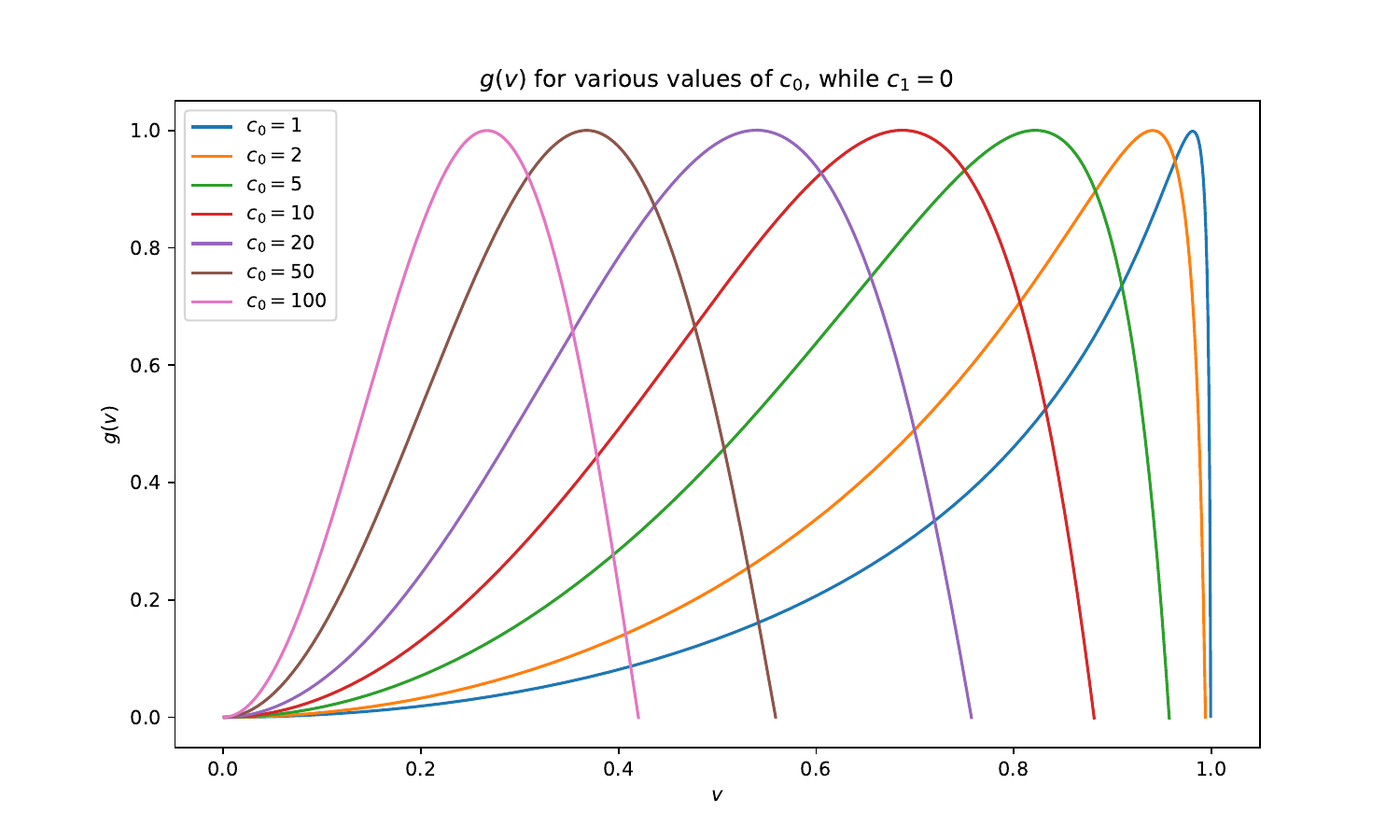}
    \caption{Z-family solutions for $g(v)$ for various values of $c_0$ when $c_1=0$.} 
    \label{fig:1}
\end{figure}

In the absence of pressure and inertia, we have $c_1=0$. The results of the integration for $c_1=0$ and various values of $c_0$, ranging from $1$ to $100$, are shown in Fig.~ \ref{fig:1}. They are fully consistent with the analytical solutions reported by \cite{Barkov:2022}, with the note that the value of $\alpha$ there corresponds to $\sqrt{c_0}$ in this work. Clearly, when the value of $c_0$ increases, the maximum velocity attained by the expanding field becomes lower. This velocity corresponds to the first root, beyond the origin $v=0$, as $g(0)=0$, of the function $g(v_0)=0$. Thus, by increasing the ratio of the poloidal current over the poloidal field, or equivalently, the ratio of the toroidal field over the poloidal field, the maximum expansion velocity is lower.
We note that beyond $v_0$, the solutions become oscillatory, with the flux taking negative and positive values. However, this part of the solution is completely disconnected from the inner part in terms of the poloidal and the toroidal fields, and thus, no field lines from the central part are connected to the outer ones. Thus, we halted the integration at $v_0$. These solutions are denoted as Z-type solutions, as the pressure and density are zero. 

\subsection{General solution}

When all terms are included so that $c_0\neq 0$ and $c_1\neq 0$ and the system is relativistic, the equation does not permit general analytical solutions, and we proceed with a numerical solution. Given the singularities at $v=0$ and $v=1$, we integrated within an interval $v\in [v_{min}, v_{max}]$, where $v_{min}=10^{-3}$ and $v_{max}<1$. We set $g(v_{min})^{\prime}=0$ and $g(v_{min})=g_0$, and we chose $g_{0}$ so that the maximum value of $g$ was set to unity. This is straightforward for the homogeneous version of the equation with $c_1=0$, as a simple division with the maximum value of a trial solution with arbitrary $g_0$ leads to a normalised solution. However, in the non-homogeneous case ($c_{1}\neq0$), it requires an iterative processes as  simple division of the function does not lead to a solution. Thus, we repeated the process by dividing $g_{0}$ by the maximum value of the solution and integrated until convergence. We used a tolerance limit of $10^{-4}$ for the normalisation, and we typically found that the solution converged to a normalised form after fewer than $100$ iterations. For the integration of the ordinary differential equation, we used the Runge-Kutta-Fehlberg method, applying the built-in routines of numpy and scipy in Python \citep{harris2020array,2020SciPy-NMeth}.

\begin{figure}
     \includegraphics[width=0.45\textwidth]{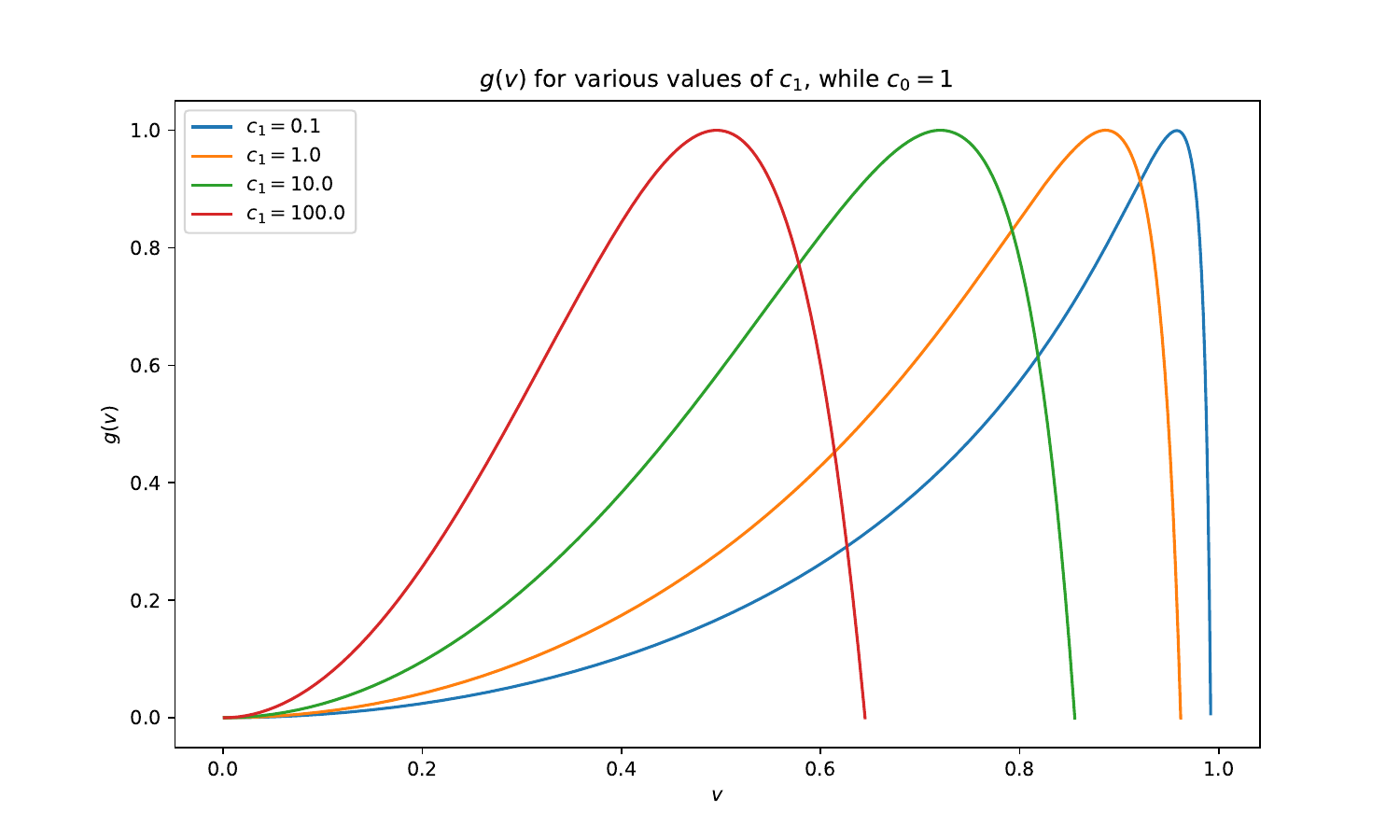}
    \caption{Characteristic P-family  solutions of $g(v)$ for various values of $c_1$ while keeping $c_0=1$. } 
    \label{fig:2}
\end{figure}

Next, we explored solutions with non-zero values of $c_1$. Because of the very many possible combinations, we decided to keep the toroidal field parameter constant ($c_0=1$) and then explore positive values of $c_1$ ranging from $0.1$ to $100$. The results are shown in Fig.~ \ref{fig:2}. The inclusion of pressure and density leads to a structure that is less extended in the velocity space and reaches a lower maximum velocity than in the $c_1=0$ case. We denoted these solutions P-type solutions, as the pressure and density increase with increasing magnetic flux function.

\begin{figure}
     \includegraphics[width=0.45\textwidth]{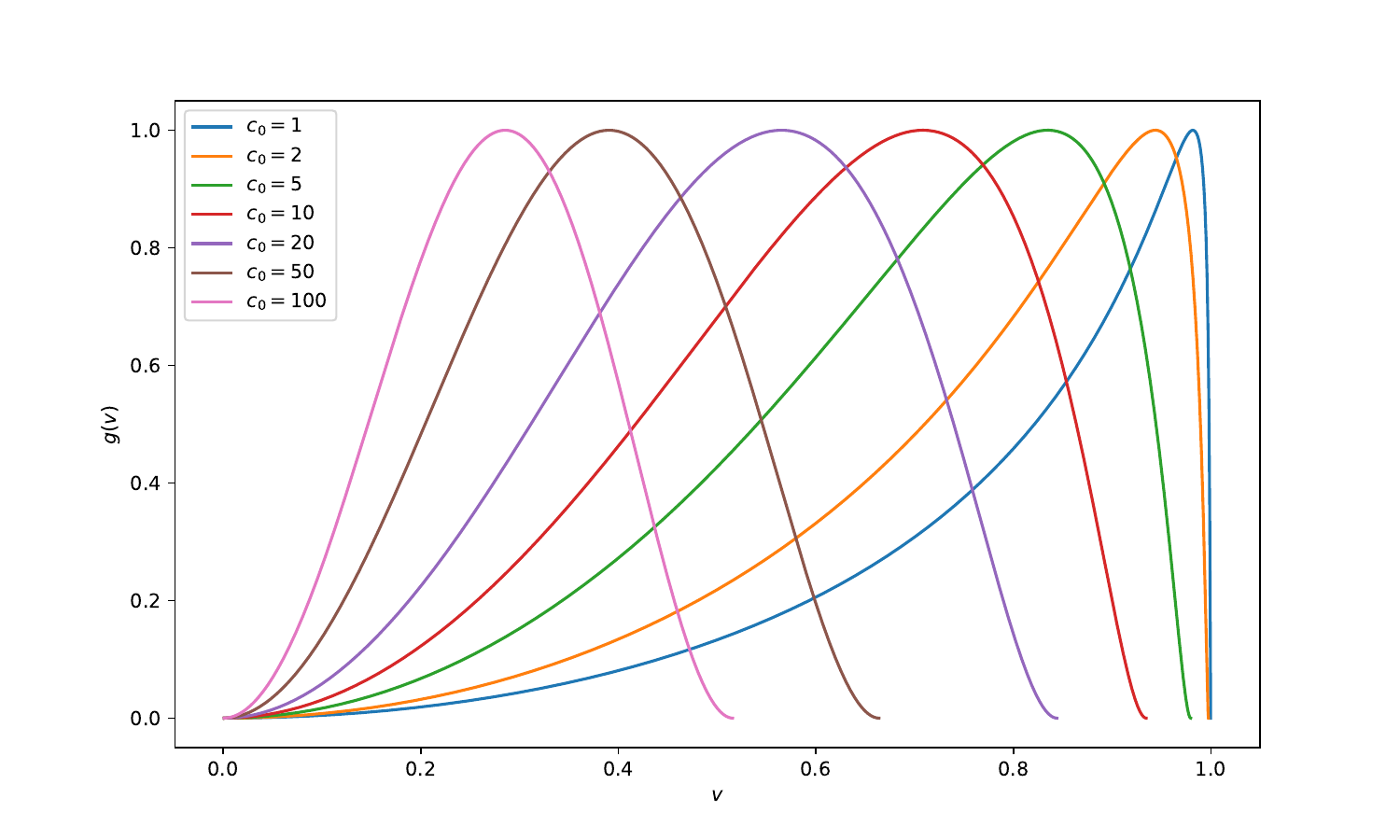}
    \caption{N-family of solutions showing $g(v)$ for various values of $c_0$ and negative $c_1$, so that the derivative is zero at the maximum velocity.  } 
    \label{fig:3}
\end{figure}

Finally, it is possible to explore negative values of $c_1$. These cases correspond to plasmoids where the pressure decreases as the magnetic flux function increases. Despite this negative relation, the freedom of an additive constant $\tilde{p}_0$ in Eq. (\ref{eq:pressure}) prevents negative values of the pressure and density. These solutions have the remarkable property that for an appropriate combination of $c_0$ and $c_1$, not only the flux function reaches $0$ at $v_{0}$, but its derivative also becomes zero there, and thus, $g^{\prime}(v_0)=0$, as shown in Fig.~ \ref{fig:3}. This has the important consequence that all three components of the magnetic field are zero on the surface of the sphere. We note that when the values of $c_1$ are smaller algebraically than those corresponding to the zero derivative on the boundary, the solutions become unphysical, with the field lines not closing, but increasing infinitely. The resulting structures correspond to underdensities, where the drop of thermal plasma pressure and mass density within the sphere is compensated for by the magnetic and electric field pressure and tension. We denoted these as N-type solutions as in these solutions the density and pressure decrease with increasing flux function.

The parameters provided in Table \ref{tab:1} were employed for the three families of solutions: $c_1=0$ Z-models, without mass or thermal pressure, $c_1>0$ P-models, with varying mass and pressure with the same sign as the flux function, and $c_1<0$ N-models, where, in addition to the negative sign of $c_1$, the derivative of $g$ at $v_0$ vanishes.
\begin{table*}[h!]
\centering
\caption{Solution properties. }
\begin{tabular}{ccccccccc}
\hline
\hline
Model & $c_0$ & $c_1$ & $v_{0}$ &$\Gamma_0$& $E_{mag}$& $E_{el}$& $W$ & $M$  \\
\hline
Z1 & 1 & 0 & 0.9992 &  25.01 & 17.59 &17.54 & 0 & 0  \\
Z2 & 2 & 0 & 0.9939 & 9.07 & 1.018 &0.9730 & 0 & 0 \\
Z3 & 5 & 0 & 0.9573 & 3.46 & 0.2691 &0.1998 & 0 & 0  \\
Z4 & 10 & 0 & 0.8812 & 2.12 & 0.2004 &0.1012 & 0 & 0  \\
Z5 & 20 & 0 & 0.7571 & 1.53 & 0.2020 &0.06043 & 0 & 0  \\
Z6 & 50 & 0 & 0.5589 & 1.21 & 0.2598 &0.03446 & 0 & 0  \\
Z7 & 100 & 0 & 0.4201 & 1.10 & 0.3431 &0.02354 & 0 & 0  \\
\hline
P1 & 1 & 0.1 & 0.9918 &  7.82& 1.742 &1.693 & 0.02989 & 0.07039  \\
P2 & 1 & 0.2 & 0.9870 & 6.22 & 1.012 &0.9656 & 0.03759 & 0.08894 \\
P3 & 1 & 0.5 & 0.9758 & 4.57 & 0.5457 &0.4988 & 0.04932 & 0.1170  \\
P4 & 1 & 1 & 0.9617 & 3.62 & 0.3654 &0.3166 & 0.05971 & 0.1406 \\
P5 & 1 & 2 & 0.9407 & 2.95 & 0.2614 &0.2094 & 0.07180 & 0.1657  \\
P6 & 1 & 5 & 0.8990 & 2.28 & 0.1872 &0.1286 & 0.09086 & 0.1990  \\
P7 & 1 & 10 & 0.8553 & 1.93 & 0.1593 &0.09345 & 0.1085 & 0.2236  \\
P8 & 1 & 20 & 0.8008 & 1.67 & 0.1445 &0.06976 & 0.1292 & 0.2458  \\
P9 & 1 & 50 & 0.7154 & 1.43 & 0.1392 &0.04921 & 0.1628 & 0.2711  \\
P10 & 1 & 100 & 0.6446 & 1.31 & 0.1435 &0.03880 & 0.1937 & 0.2865  \\
\hline
N1 & 1 & -0.00135 & 0.9997 &  40.83 & 0.8867 &0.8641 & 0.03083 & 0.02894  \\
N2 & 2 & -0.00170 & 0.9974 & 13.88 & 0.5897 &0.5504 & 0.03353 & 0.05305 \\
N3 & 5 & -0.252 & 0.9779 & 4.78 & 0.2631 &0.1970 & 0.05370 & 0.1107  \\
N4 & 10 & -1.448 & 0.9342 & 2.80 & 0.2024 &0.1065 & 0.07718 & 0.1654  \\
N5 & 20 & -7.084 & 0.8434 & 1.86 & 0.2035 &0.06541 & 0.1108 & 0.2194  \\
N6 & 50 & -50.45 & 0.6628 & 1.34 & 0.2592 &0.03796 &  0.1767 & 0.2732  \\
N7 & 100 & -211.3 & 0.5135 & 1.17 & 0.3407 &0.02609 & 0.2509 & 0.2983  \\
\hline
\hline
\label{tab:1}
\end{tabular}
\tablefoot{The columns represent the solution number, parameters $c_0$ and $c_1$, the root of the solution $v_0$, and the corresponding Lorentz number $\Gamma_0$, the magnetic field energy $E_{mag}$, the electric field energy $E_{el}$, the work of the pressure $W$, and the mass $M$ enclosed in the solution. The units of the various energy terms ($E_{mag}$, $E_{el}$, and $W$) are in units of $10^{45}$ erg and the mass is in units of $1.8\times 10^{26}$g, under the assumption of a sphere where $v=1$ corresponds to a radius of $1$km, and the magnetic flux $g$ is normalised to $10^{25}$G~cm$^2$.  }
\end{table*}

\section{Discussion}
\label{sec:discussion}

\subsection{General properties of the solutions}

We describe the main characteristics of the solutions depending on the values of the parameters below. For all solutions, lower values of $c_0$ lead to a $v_0$ that approaches unity, and thus, the system becomes more relativistic. This implies that the more twisted the solution, or equivalently, the higher the poloidal current, the smaller the sphere containing the plasmoid. This property is related to the fact that by assuming a constant $c_0$, a particular length-scale is introduced. In this case, the units of $\sqrt{c_0}$ are inversely proportional to the length. Consequently, by increasing $c_0$, this length scale becomes smaller. This is  consistent with the well-known non-relativistic linear force-free axisymmetric solutions in spherical geometry \citep{Chandrasekhar:1957}, whose radius shifts to lower values for a higher toroidal-to-poloidal magnetic field ratio. This property is well preserved and reflected on its relativistic analogue.

In the Z and P solutions, a non-vanishing meridional magnetic field exists on the surface $v=v_0$. Because of this magnetic field discontinuity, an azimuthal electric current sheet flows on the surface of the plasmoid. Moreover, there also exists a volume electric charge because the meridional component of the electric field diverges. Following the convention of a positive value for the flux function, the charge is positive in the northern and negative in the southern hemisphere. While this structure is in equilibrium at its interior, the magnetic pressure is anisotropic on its surface, stronger at the equator, and weaker at the poles. Therefore, when it is embedded within a medium of uniform pressure, it is expected to expand in the equatorial direction, while the surface current might trigger resistive instabilities and lead to reconnection in a layer that propagates relativistically.

Considering the solutions with $c_1>0$, we find that higher values of $c_1$ lead to plasmoids with smaller $v_0$. This is due to the inclusion of additional pressure and density that prevent the system from extending to a $v_0$ as high as in the purely electromagnetic case ($c_1=0$). The characteristics of the overall structure are similar to those of the previous solution regarding the surface azimuthal current sheet and pressure profile, and the structure is prone to similar instabilities. 

Solutions of the N family with $c_1<0$ lead to different qualitative properties. In general, choosing a negative $c_1$ leads to $v_0$ closer to unity than in solutions with zero or positive $c_1$ and the same $c_0$. Below a critical value of $c_1$, the solution has no roots and thus becomes unphysical. In the N-class of solutions, we determined these critical values for various choices of $c_0$. The solutions corresponding to these critical values have the remarkable property that at $v_0$, not only is the function $g$ zero, but its derivative also vanishes. This implies that all three components of the magnetic field and the two components of the electric field are zero on the spherical surface defined by $v=v_0$. Because of this, there is no anisotropy or a surface current due to the meridional component. Thus, when spherical plasmoids obeying these solutions are embedded within a medium of uniform pressure, the system can maintain its equilibrium, as opposed to structures corresponding to the other families of solutions. Moreover, the lack of surface current sheets makes the appearance of current-driven instabilities less likely.

\subsection{Physical quantities}

\begin{figure*}
     a\includegraphics[height=0.225\textheight]{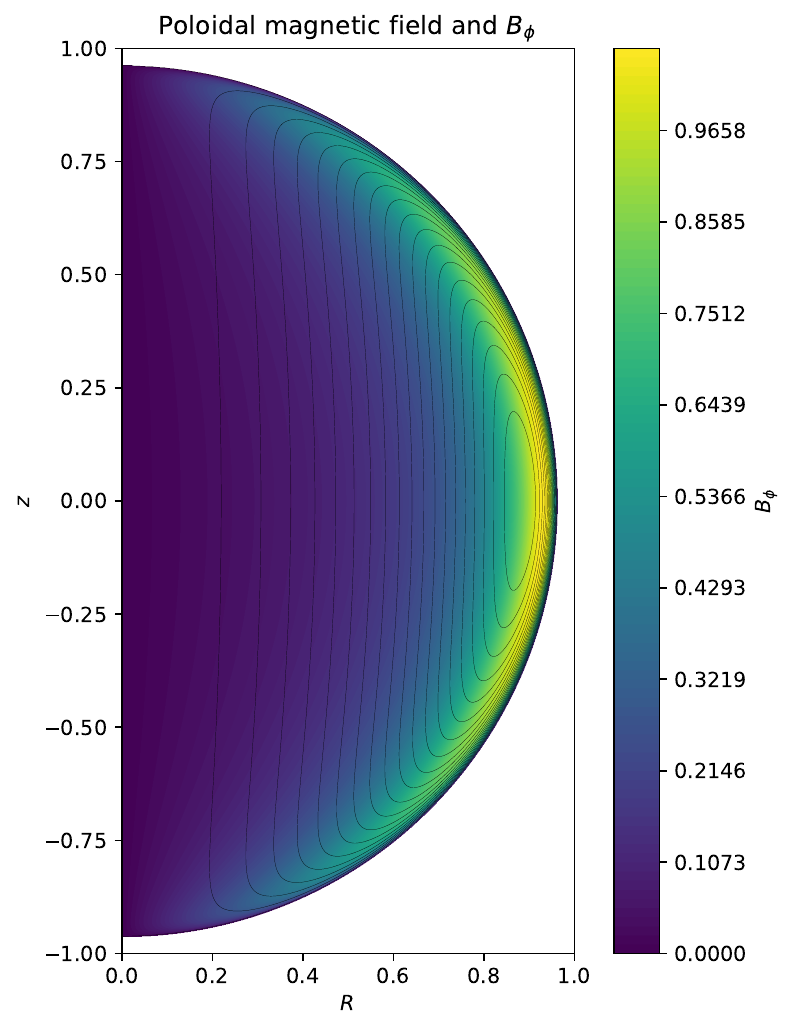}
    b\includegraphics[height=0.225\textheight]{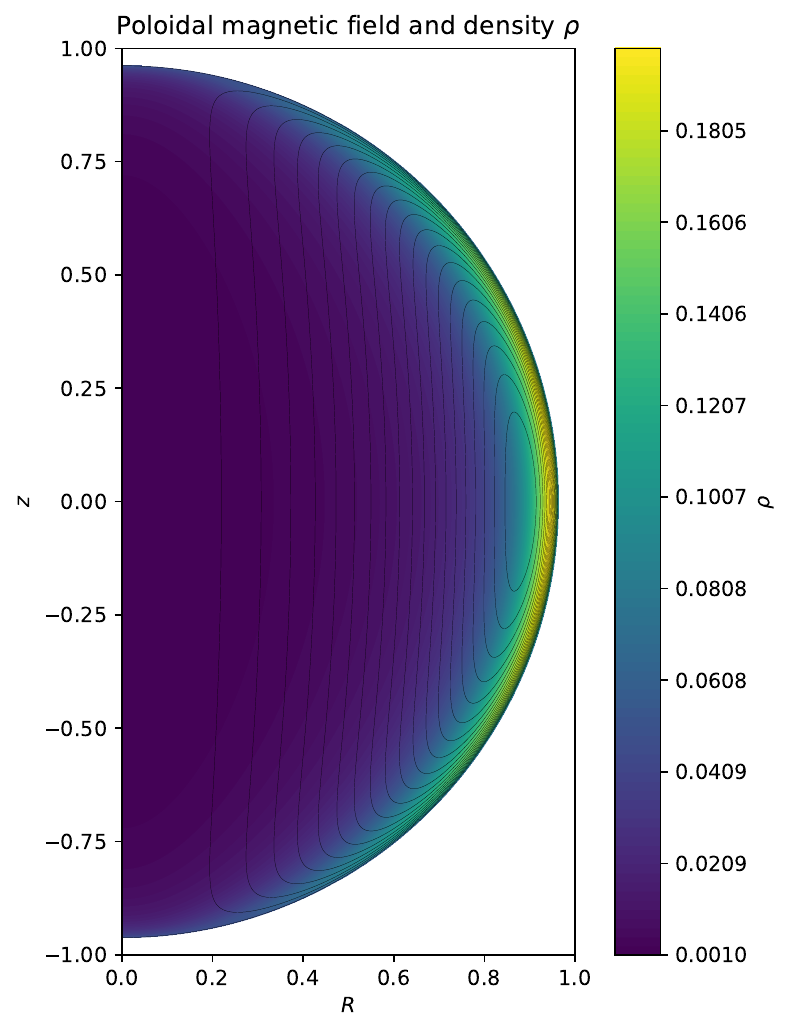}
    c\includegraphics[height=0.225\textheight]{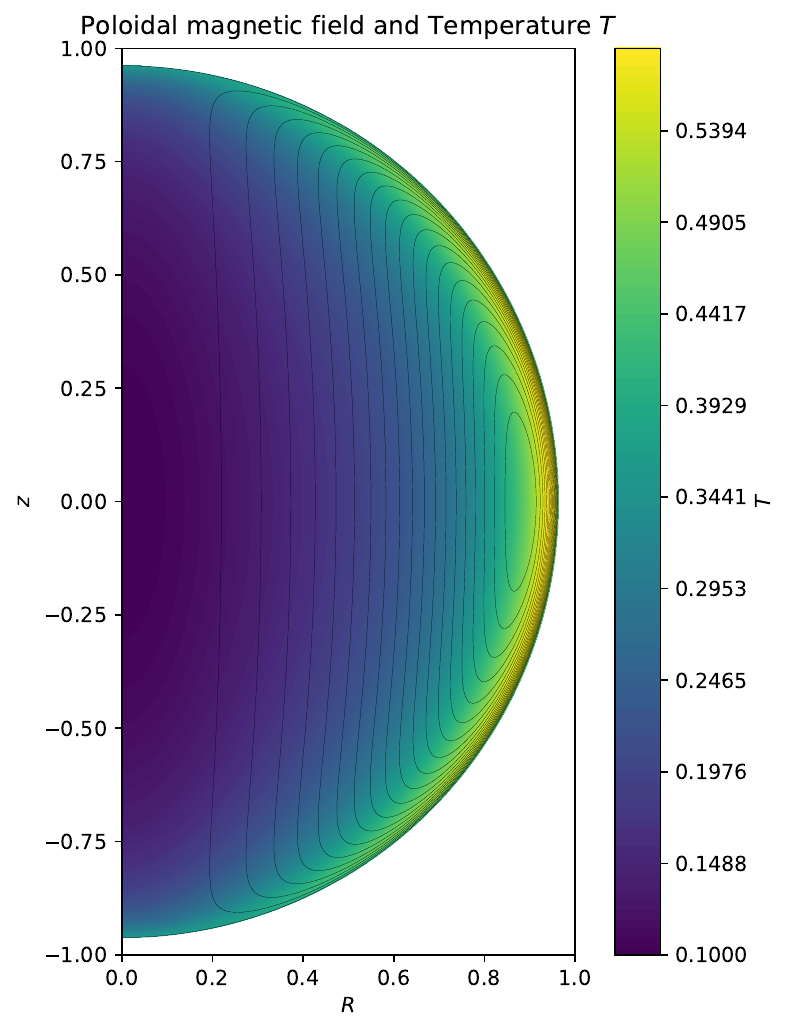}
    d\includegraphics[height=0.225\textheight]{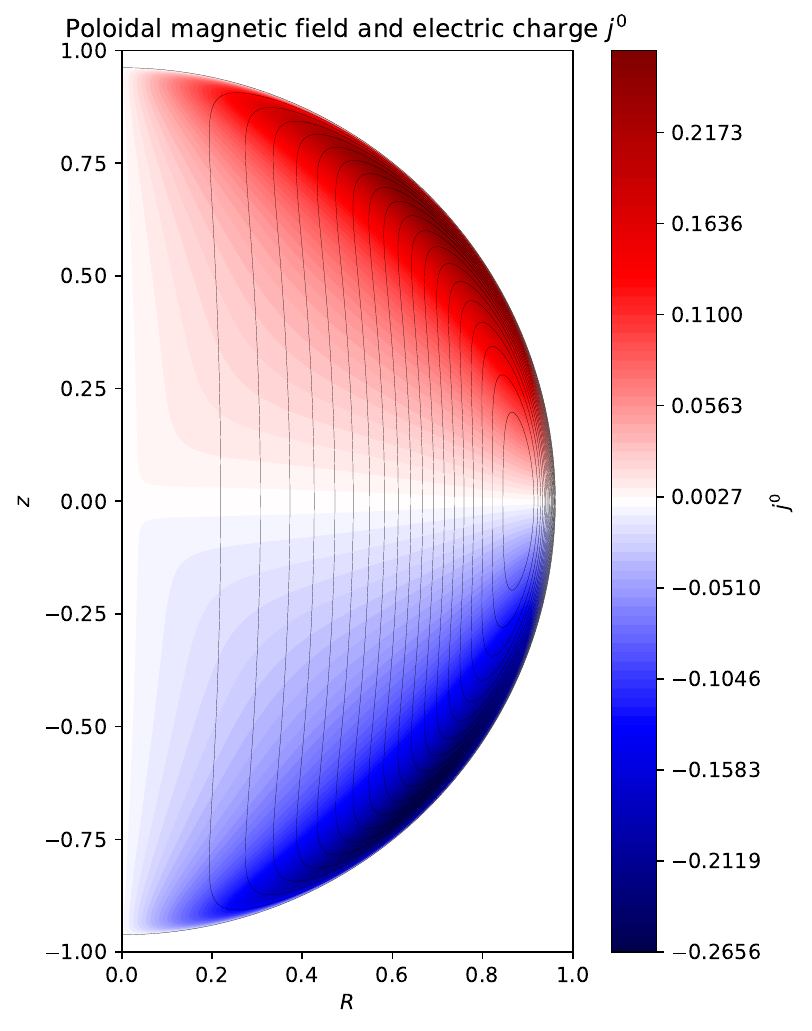}
    e\includegraphics[height=0.225\textheight]{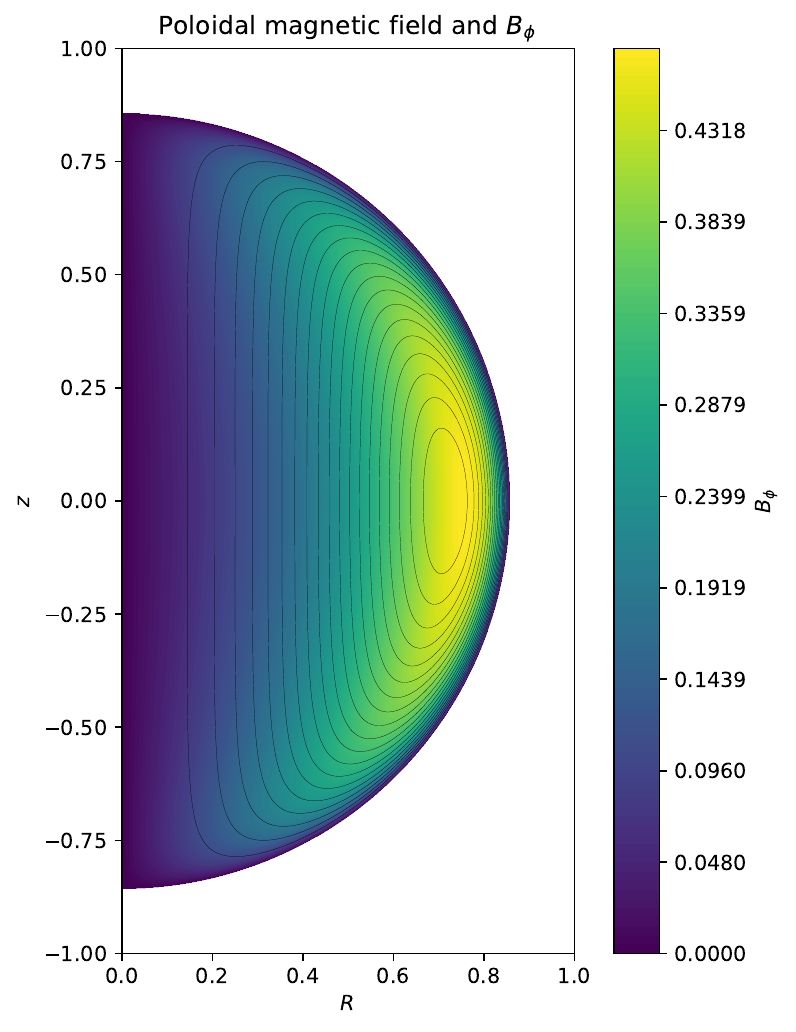}
    f\includegraphics[height=0.225\textheight]{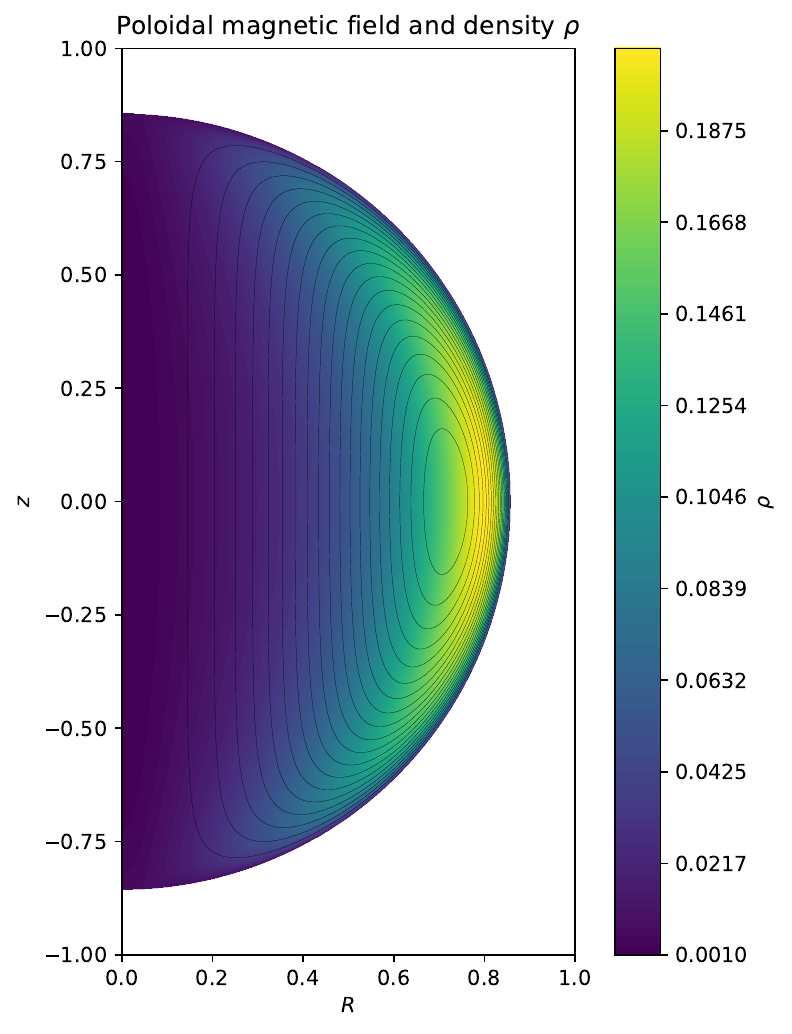}
    g\includegraphics[height=0.225\textheight]{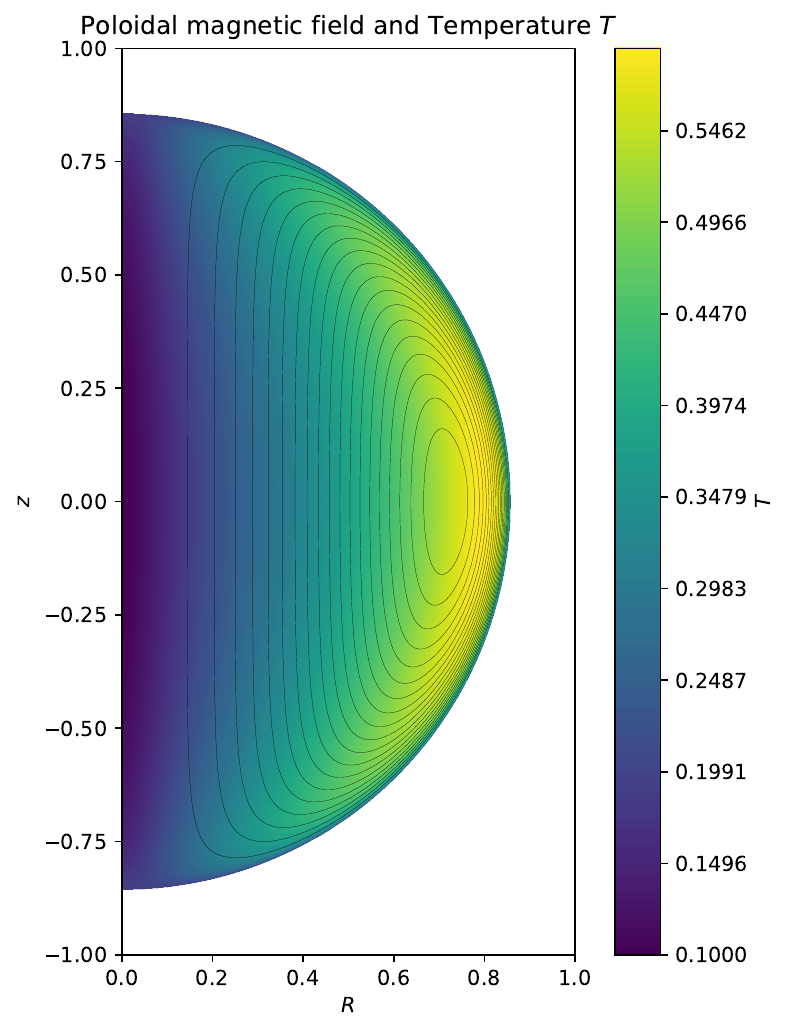}
    h\includegraphics[height=0.225\textheight]{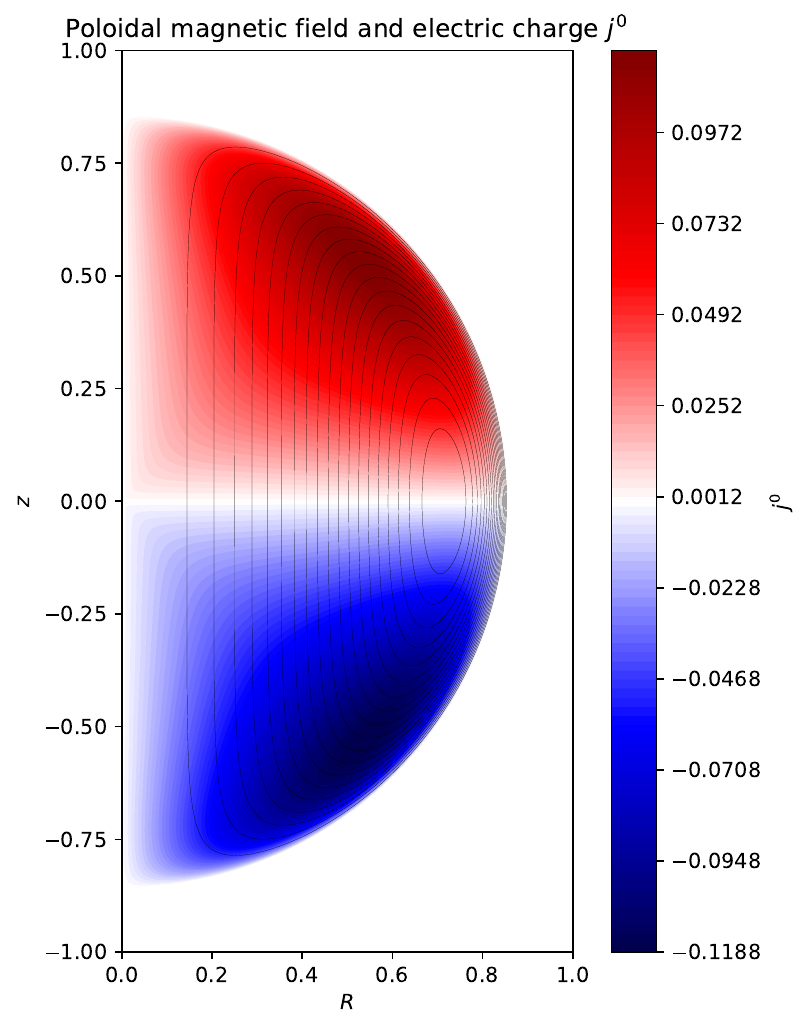}
    \caption{Contour plots of the solutions with $c_0=1$ and $c_1=1$ (top row) and $c_0=1$ and $c_1=10$ (bottom row). The black contours correspond to the poloidal magnetic field lines, and we plot in colour the $B_{\phi}$ component of the magnetic field (panels a and e), the rest-mass density $\rho$ (panels b and f), the temperature $T$ (panels c and g), and the electric charge density $j^0$ (panels d and h). The solution remains invariant in the $v$-space. We present a snapshot at some time $t$, so that $r=1$ corresponds to $v=1$.  } 
    \label{fig:Panel1}
\end{figure*}

\begin{figure*}
     a\includegraphics[height=0.225\textheight]{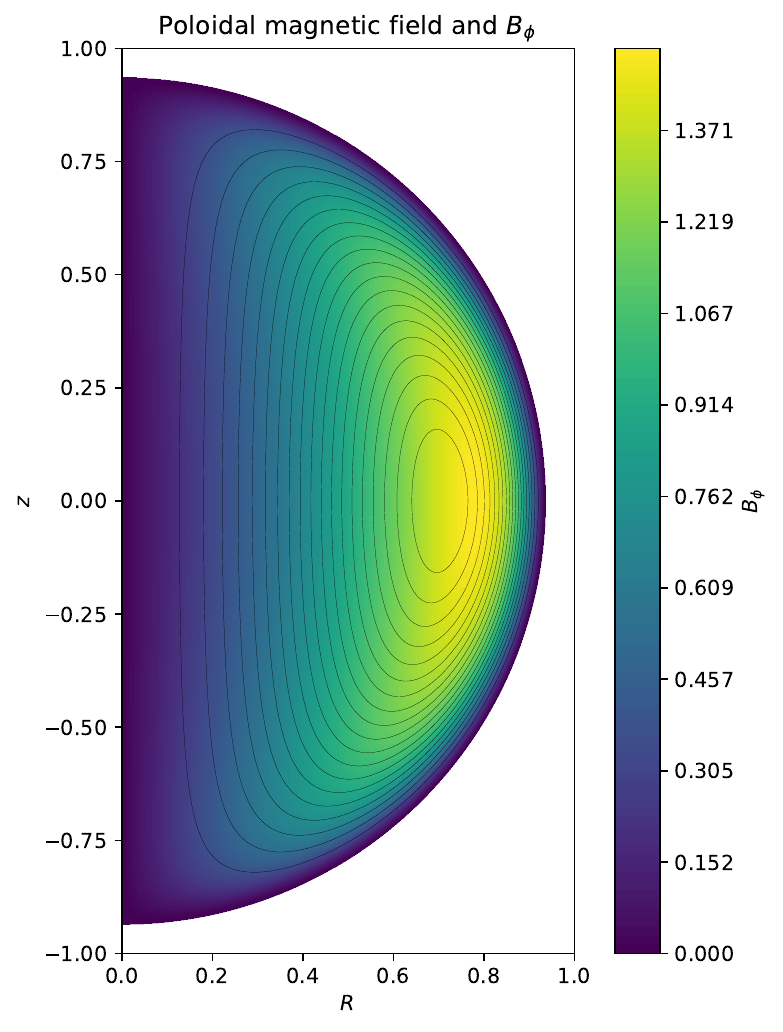}
    b\includegraphics[height=0.225\textheight]{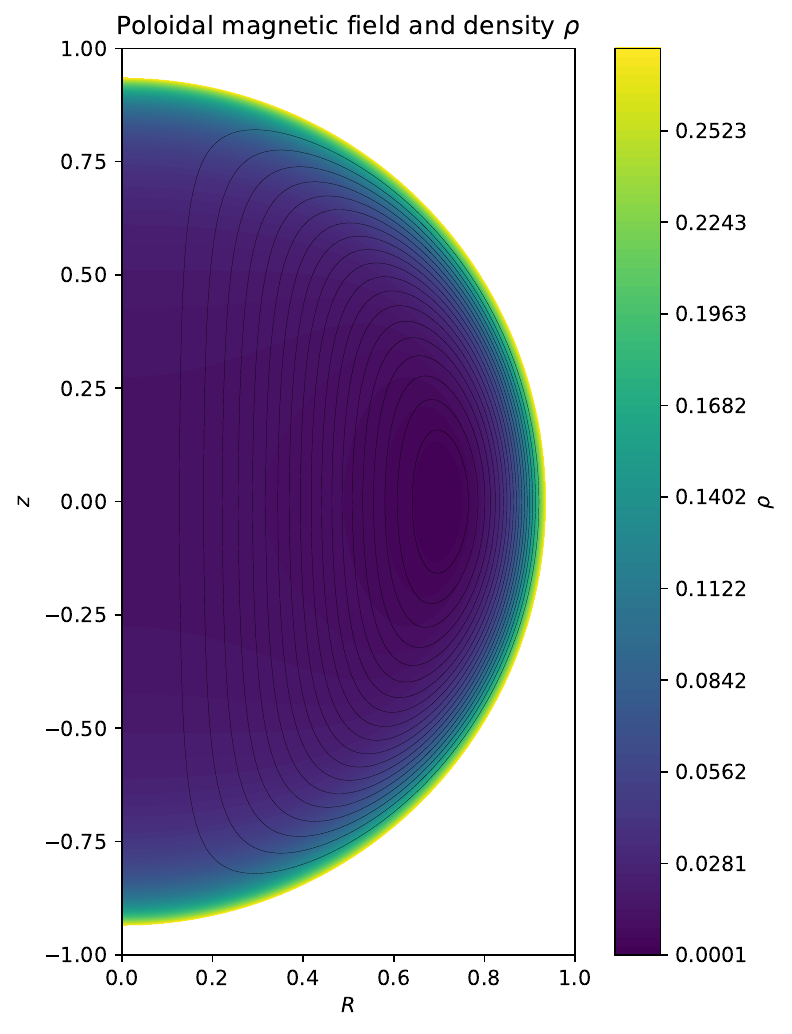}
    c\includegraphics[height=0.225\textheight]{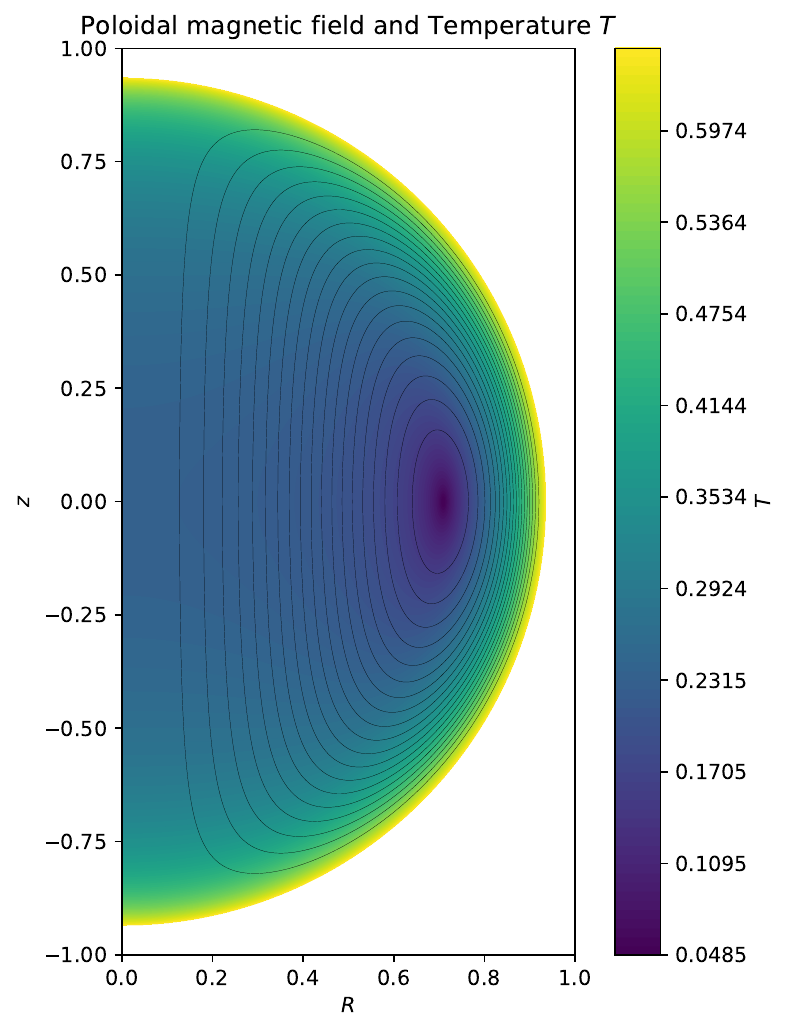}
    d\includegraphics[height=0.225\textheight]{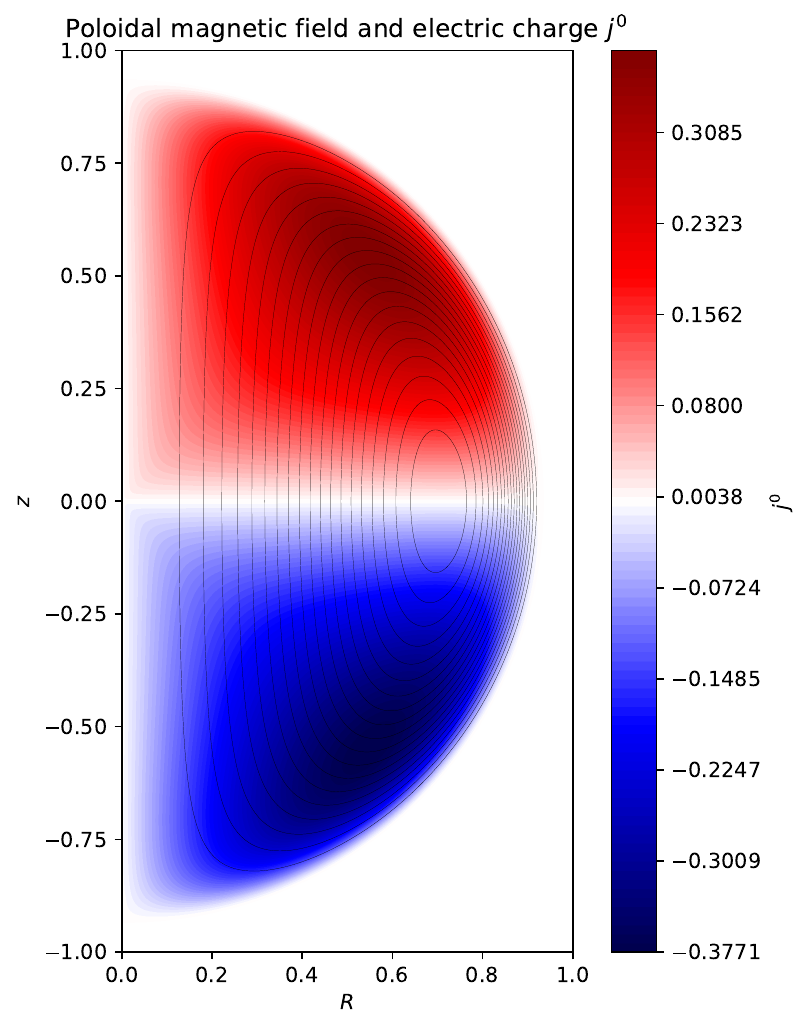}
    e\includegraphics[height=0.225\textheight]{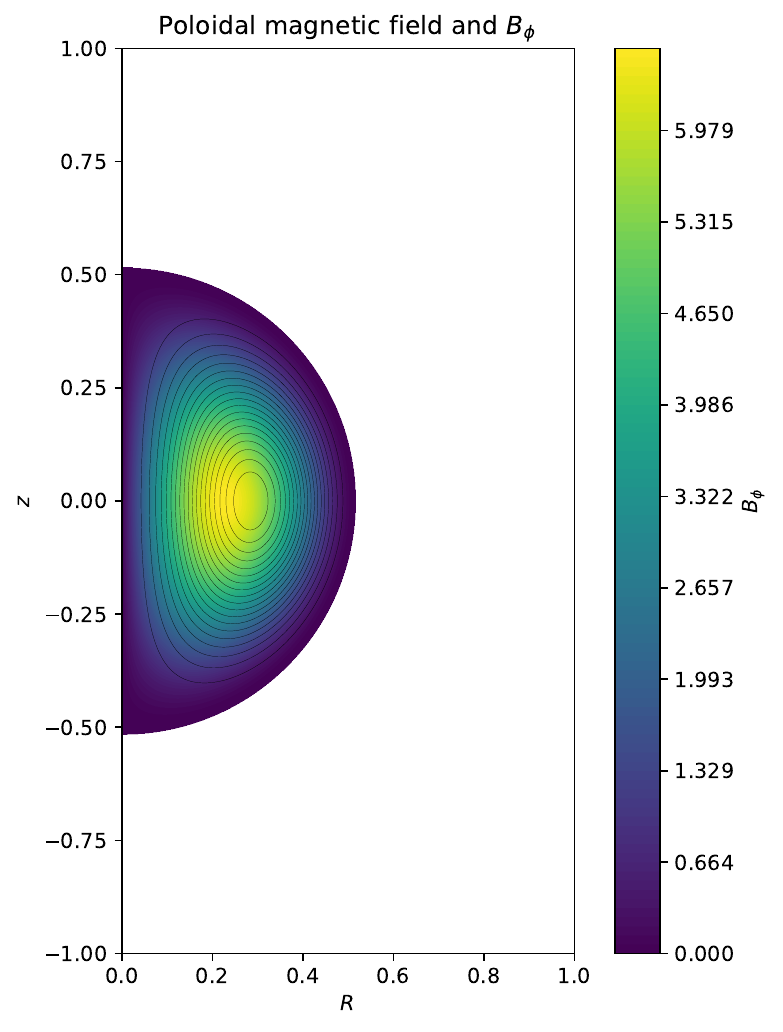}
    f\includegraphics[height=0.225\textheight]{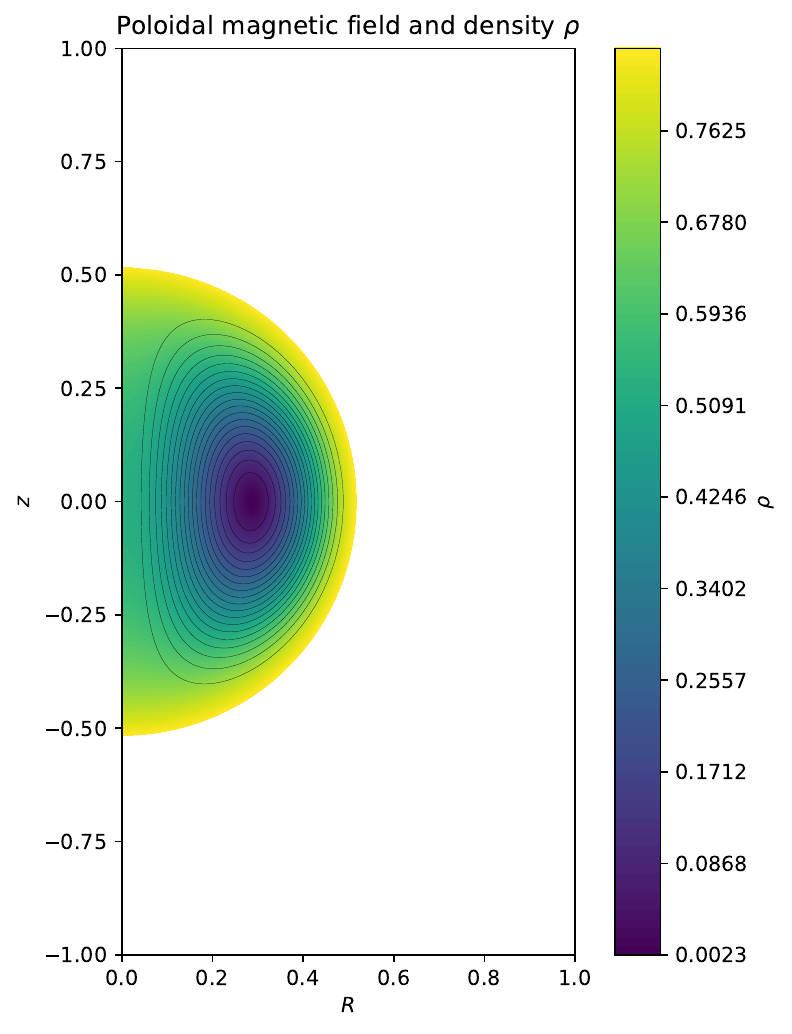}
    g\includegraphics[height=0.225\textheight]{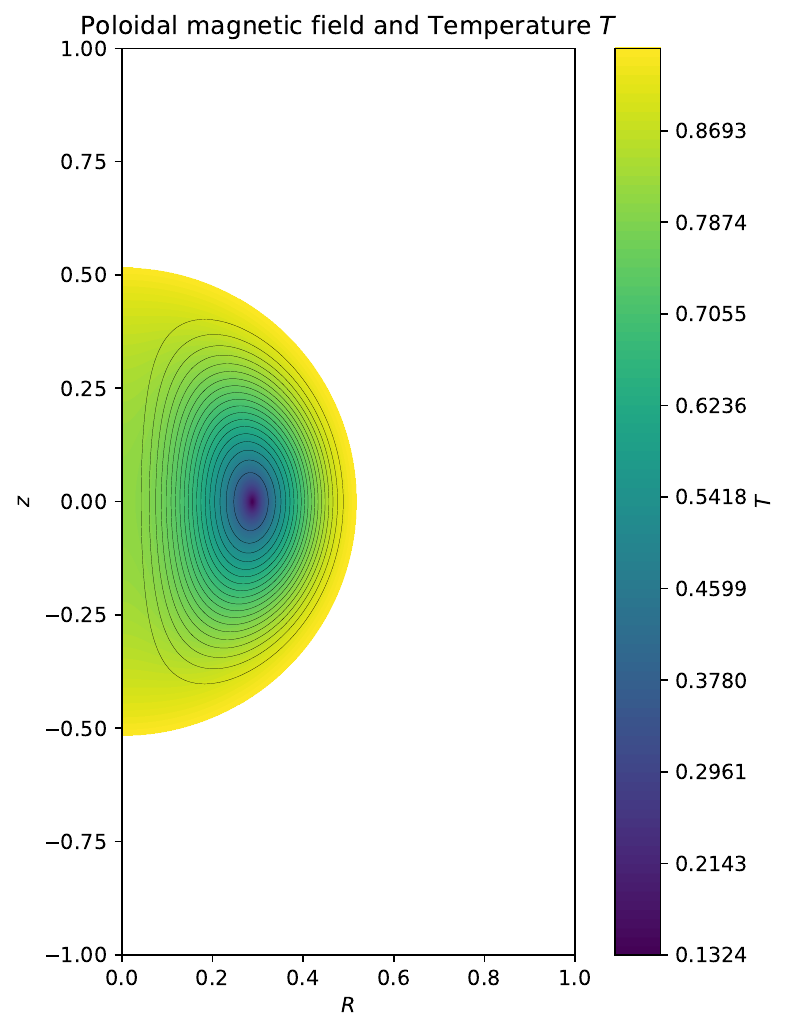}
    h\includegraphics[height=0.225\textheight]{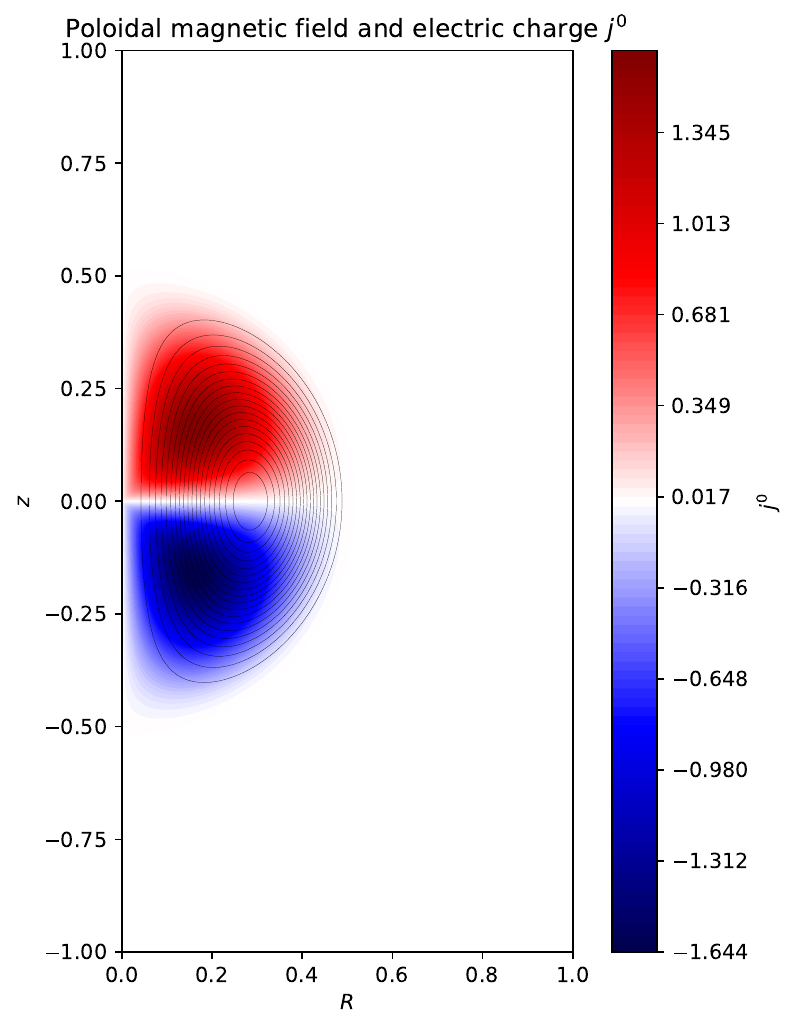}
    \caption{Contour plots of the solutions with $c_0=10$ and $c_1=-1.448$ (top row) and $c_0=100$ and $c_1=-211.3$ (bottom row). The black contours correspond to the poloidal magnetic field lines, and we plot in colour the $B_{\phi}$ component of the magnetic field (panels a and e), the rest-mass density $\rho$ (panels b and f), the temperature $T$ (panels c and g), and the electric charge density $j^0$ (panels d and h). As in the plots of Fig.~\ref{fig:Panel2}, the snapshot is taken at some time $t$, so that $r=1$ corresponds to $v=1$. } 
    \label{fig:Panel2}
\end{figure*}

Based on these solutions, it is possible to calculate the corresponding physical quantities arising from these calculations. In what follows, we focus on the electromagnetic field energy, the mass of the sphere, the work of the thermal pressure, and the temperature of the plasmoid. A set of characteristic configurations and the corresponding  physical properties are shown in Figs.~ \ref{fig:Panel1} and \ref{fig:Panel2}.

\subsubsection{Electromagnetic field}

The electric and magnetic fields are given from Eqs. (\ref{eq:Ohm}) and  (\ref{eq:mag}) with appropriate substitution of the solutions and the velocity profile. The electromagnetic field is contained within a sphere in the $v$-space. The poloidal field is approximately uniform close to the axis, and it closes within an expanding sphere of radius $R=cv_0 ~t$. The $B_{\phi}$ field peaks at the equatorial plane. However, this peak does not coincide with the maximum of the flux function $\Psi$, which  corresponds to the null point of the closed poloidal loops (Figs.~\ref{fig:Panel1} and \ref{fig:Panel2} panels a and e) because $B_{\phi}$ is proportional to $\Psi$, but is also multiplied by a factor $\frac{v\Gamma^2}{r^2}$, which for the highly relativistic case dominates and pushes the maximum to more distant radii. Therefore, in the highly relativistic case, a thin layer containing a toroidal field expanding relativistically is expected, while in the non-relativistic case $v_0\ll 1$, the toroidal field is contained closer to the centre.  

We then calculated the energy of the electric and the magnetic field,
\begin{eqnarray}
    E_{mag} &=&\frac{2\pi}{8 \pi} \int_0^{\pi}\int_0^{R}\frac{1}{(2 \pi r^2 \sin{\theta})^2}\Big[\left(\partial_\theta \Psi\right)^2+  \nonumber\\
    &+&\left(v\partial_v \Psi\right)^2+\left(\Gamma^2 v \sqrt{c_0} \Psi\right)^2\Big]r^2\sin\theta dr d\theta \nonumber \\
    &=&\frac{1}{16 \pi ct}\int_0^{\pi} \int_0^{v_0}\frac{1}{v^2 \sin^2\theta }\Big[\left(\partial_\theta \Psi\right)^2+  \nonumber\\
    &+&\left(v\partial_v \Psi\right)^2+\left(\Gamma^2 v \sqrt{c_0} \Psi\right)^2\Big]\sin\theta dv d\theta\,.
    \label{eq:Emag}
\end{eqnarray}
It follows from Eq. (\ref{eq:Emag}) that the magnetic field energy scales inversely with time. 
This behaviour can be understood from simple scaling arguments: under self-similar expansion, the magnetic flux is conserved, implying that the characteristic magnetic field strength scales as \( B \propto R^{-2} \). 
Since the magnetic energy density scales as \( B^2 \propto R^{-4} \), integration over the expanding spherical volume (\( \propto R^3 \)) yields a total magnetic energy \( E_{\rm mag} \propto R^{-1} \). 
Because the outer radius of the configuration evolves as \( R = c v_0 t \), this directly leads to \( E_{\rm mag} \propto t^{-1} \).

\begin{figure}
     \includegraphics[width=0.45\textwidth]{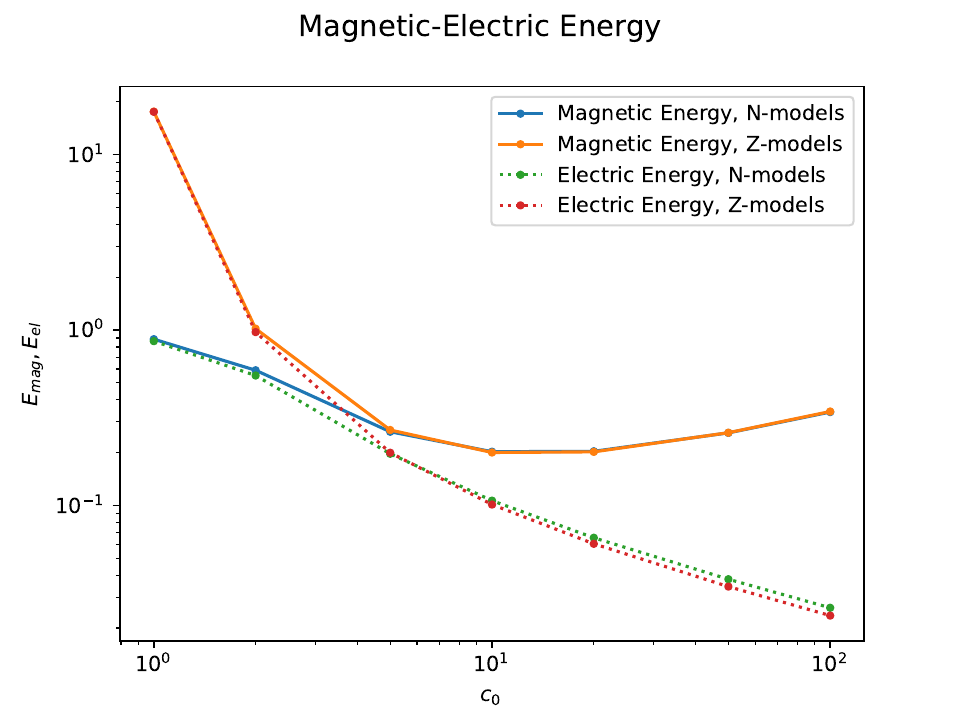}
    \caption{Electric and magnetic energy for the Z and N solutions as functions of $c_0$. } 
    \label{fig:electric_magnetic_energy}
\end{figure}

Similarly, the energy stored in the electric field is given from the following expression:
\begin{eqnarray}
    E_{el} &=&\frac{2\pi}{8 \pi} \int_0^{\pi}\int_0^{R}\frac{v^2}{(2 \pi r^2 \sin{\theta})^2}\Big[ \left(v\partial_v \Psi\right)^2+\left(\Gamma^2 v \sqrt{c_0} \Psi\right)^2\Big]\times \nonumber \\
    &\times &r^2\sin\theta dr d\theta \nonumber \\
    &=&\frac{1}{16 \pi ct}\int_0^{\pi} \int_0^{v_0}\frac{1}{\sin^2\theta}\Big[\left(v\partial_v \Psi\right)^2+  \nonumber\\
    &+&\left(\Gamma^2 v \sqrt{c_0} \Psi\right)^2\Big]\sin\theta dv d\theta\,.
\end{eqnarray}
As the electric field is the $\theta$ and $\phi$ components of the magnetic field multiplied by the velocity, it follows a similar behaviour, with the electric field energy being inversely proportional to time. Thus, as the system expands, it decreases. We show the electric and magnetic field energy contained within a sphere as a function of $c_0$ in Fig. \ref{fig:electric_magnetic_energy}, where the solution for the Z ($c_1=0$) and N-models ($c_1<0)$ is shown. For lower and higher values of $c_0$, the magnetic field energy is higher and has a minimum for intermediate $c_0$. We remark that we normalised the maximum value of $g$ and consequently $\Psi$ to unity in the solutions. Because of this, while the magnetic flux is the same in all solutions, its energy content is related to its distribution in the sphere. In the case of a smaller sphere and large $c_0$, all components of the magnetic field are large since they scale inversely with $r$. For spheres where $v_0$ approaches unity, while the radius is higher, $g$ develops a sharp derivative at $v_0$, leading to a strong $B_{\theta}$ component and energy content. In contrast, the electric field energy has a monotonic dependence on $c_0$ as its intensity is proportional to $v$, and therefore, its overall contribution is smaller for smaller spheres in velocity space, even though the magnetic field energy is higher. 

\subsubsection{Mass}

The density profile of the plasmoid is related to the flux function $\Psi$, Eqs. (\ref{eq:Q}) and (\ref{eq:pressure}), multiplied by the prefactor $\Gamma^3 v^3/r^3$, as demonstrated in Eq. (\ref{eq:density}). The two families of solutions we studied here for positive and negative $c_1$ have qualitatively different density profiles. For $c_1>0$, the minimum is located at the centre of the sphere ($r=0$) and peaks close to the location of the maximum of $\Psi$ with some modulation due to the $\Gamma^3$ prefactor that pushes the maximum closer to the surface. In contrast, for $c_1<0$, the minimum is located close to the maximum of $\Psi$.

\begin{figure}
     \includegraphics[width=0.45\textwidth]{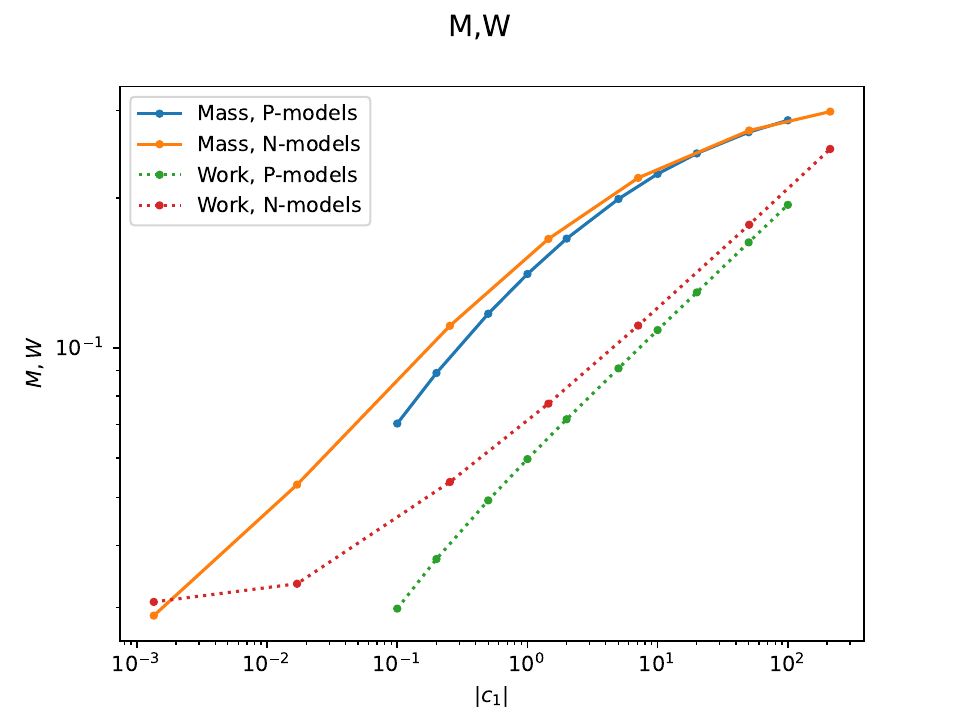}
    \caption{Mass $M$ and work $W$ for the P and N solutions as functions of $|c_1|$. While $M$ and $W$ are plotted on the same axis, the units differ. Moreover, $W$ evolves with time, which is inversely  proportional to $t$.} 
    \label{fig:mass-pressure}
\end{figure}

We first considered the rest mass contained within the expanding plasmoid. We integrated the rest-mass density expression given by Eq. (\ref{eq:density}) within a sphere of given radius $R$, so that $R=cv_{0}~t$ at some time $t$, which corresponds to the physical location of the non-zero root of $g$. By accounting for the axial symmetry of the system, integration over $\phi$ provides a factor of $2\pi$,
\begin{eqnarray}
    M&=&2\pi\int_{0}^{\pi}\!\!\int_{0}^{R}\!\left(\tilde{p}_{0}+c_1\frac{g(v)\sin^2\theta}{16 \pi^3}\right)^{3/4} Q(v,\theta)^{-3/4}\times \nonumber \\
    &&\left(\frac{\Gamma v}{r}\right)^3 r^2\sin\theta ~dr~d\theta\,\nonumber \\
    &=& 2\pi\int_{0}^{\pi}\!\!\int_{0}^{R/(ct)}\!\left(\tilde{p}_{0}+c_1\frac{g(v)\sin^2\theta}{16 \pi^3}\right)^{3/4}Q(v,\theta)^{-3/4}\times \nonumber \\
    &&\Gamma^3 \left(\frac{r}{ct}\right)^2\sin\theta  ~d\left(\frac{r}{ct}\right)~d\theta\,\nonumber \\
    &=& 2\pi\int_{0}^{\pi}\!\!\int_{0}^{v_0}\!\left(\tilde{p}_{0}+c_1\frac{g(v)\sin^2\theta}{16 \pi^3}\right)^{3/4}Q(v,\theta)^{-3/4}\times \nonumber \\
    &&\Gamma^3 v^2\sin\theta  ~dv~d\theta\,.
\end{eqnarray}
We confirm, as expected, that the rest mass contained within a spherical configuration does not evolve with time or the physical radius of the fluid, as the integral is on the $v$-space, where the solution does not evolve. However, the amount of mass contained inside the sphere depends on the solution. In what follows, we assume that the function $Q$ is constant and sets the scale for the mass. Furthermore, the total mass is sensitive to the choice of the additive constant, $\tilde{p}_0$. We made the following choices: for $c_1>0$, we set $\tilde{p}_0=0$, whereas for $c_1<0$, we chose the minimum acceptable value of $\tilde{p}_0$ so that the minimum of the density is exactly zero, and thus, $\tilde{p}_0=-\frac{c_1}{16\pi^3}$. The resulting masses from the integration are shown in Fig.~\ref{fig:mass-pressure}. We note that for higher absolute values of $c_1$ the mass is higher. We note that the assumption of an additive constant affects the direct comparison of the mass: the P-type solutions have zero density on their surface, where the N-type solutions have a non-zero density on their surfaces. To produce a reasonable comparison, the P-type solutions have a higher mass load than the N-type solutions for the same density on the surface $v=v_0$. 

\subsubsection{Thermal pressure work}

Following a similar procedure as for the density, we evaluated the work $W$ of the thermal pressure. It can be calculated through the integral of $pdV$. Upon integration for a sphere of radius $R$ at time $t$, we obtain the following expression:
\begin{eqnarray}
    W&=&2\pi\int_{0}^{\pi}\!\!\int_{0}^{R}\!\left(\tilde{p}_{0}+c_1\frac{g(v)\sin^2\theta}{16 \pi^3}\right)\times \nonumber \\
    &&\left(\frac{\Gamma v}{r}\right)^4 r^2\sin\theta ~dr~d\theta\,\nonumber \\
    &=& \frac{2\pi}{ct}\int_{0}^{\pi}\!\!\int_{0}^{v_0}\!\left(\tilde{p}_{0}+c_1\frac{g(v)\sin^2\theta}{16 \pi^3}\right)\times \nonumber \\
    &&\Gamma^4 v^2\sin\theta  ~dv~d\theta\,.
\end{eqnarray}
We note that $W$ is inversely proportional to $t$, signifying the expansion of the system. This work is associated with the spatial redistribution of the plasma mass and pressure as the system expands. We further note that the thermal pressure on the surface of the sphere decreases with time, and thus, as the system expands, less work in done against the external medium. The results of this integration for various choices is shown in Fig.~ \ref{fig:mass-pressure} with dotted lines.

\subsubsection{Temperature}

Given the relation between the density and the pressure, we evaluated the temperature of the plasmoid through the relation $T=\frac{m~ p}{k_B \rho }$, where $m$ is the particle mass, and $k_B$ is the Boltzmann constant. Therefore, the temperature is given by the form:
\begin{eqnarray}
    T= \frac{m}{k_B}\frac{\tilde{p}}{\tilde{\rho}}\frac{\Gamma v}{r}=\frac{m}{k_B}\frac{\tilde{p}}{\tilde{\rho}}\frac{\Gamma}{ct}.
\end{eqnarray}
Therefore, the local temperature is inversely proportional to $t$, with the expansion of the system leading to cooling. 

As $\tilde{p}$ and $\tilde{\rho}$ are related to each other through $Q$, we followed the assumption of a constant value of $Q(v,\theta)$ in the evaluation of the temperature, as we already chose in the calculation of mass through the density. Therefore, we expect a temperature profile for the spherical structure that decreases inversely proportionally with time. 

For positive values of $c_1$, the maximum value of the temperature approximately coincides with the maximum of the flux function $\Psi$, while it is cooler at the boundaries of the sphere and the axis (Fig.~ \ref{fig:Panel1} panels c and g). For negative values of $c_1$, the minimum of the temperature occurs close to the maximum of $\Psi$, while the surroundings are hotter, as illustrated in Fig.~ \ref{fig:Panel2} panels c and g. For the normalisation values assumed here, we refer to the caption of Table \ref{tab:1}, and when we assume a proton-electron plasma, the units of the temperature correspond to $10^{11}$K. 


\section{Astrophysical implications}
\label{sec:applications}

The solutions derived here provide a framework for the study of astrophysical systems of relativistically expanding configurations of magnetised plasma. It has been proposed that these structures might arise from magnetars expelling magnetic plasmoids, which are associated with magnetar flares \citep{Thompson:2001,Lyutikov:2006}, and possibly to FRBs \citep{Bransgrove:2025}. 

\subsection{Magnetar flares}

\begin{figure}
     \includegraphics[width=0.45\textwidth]{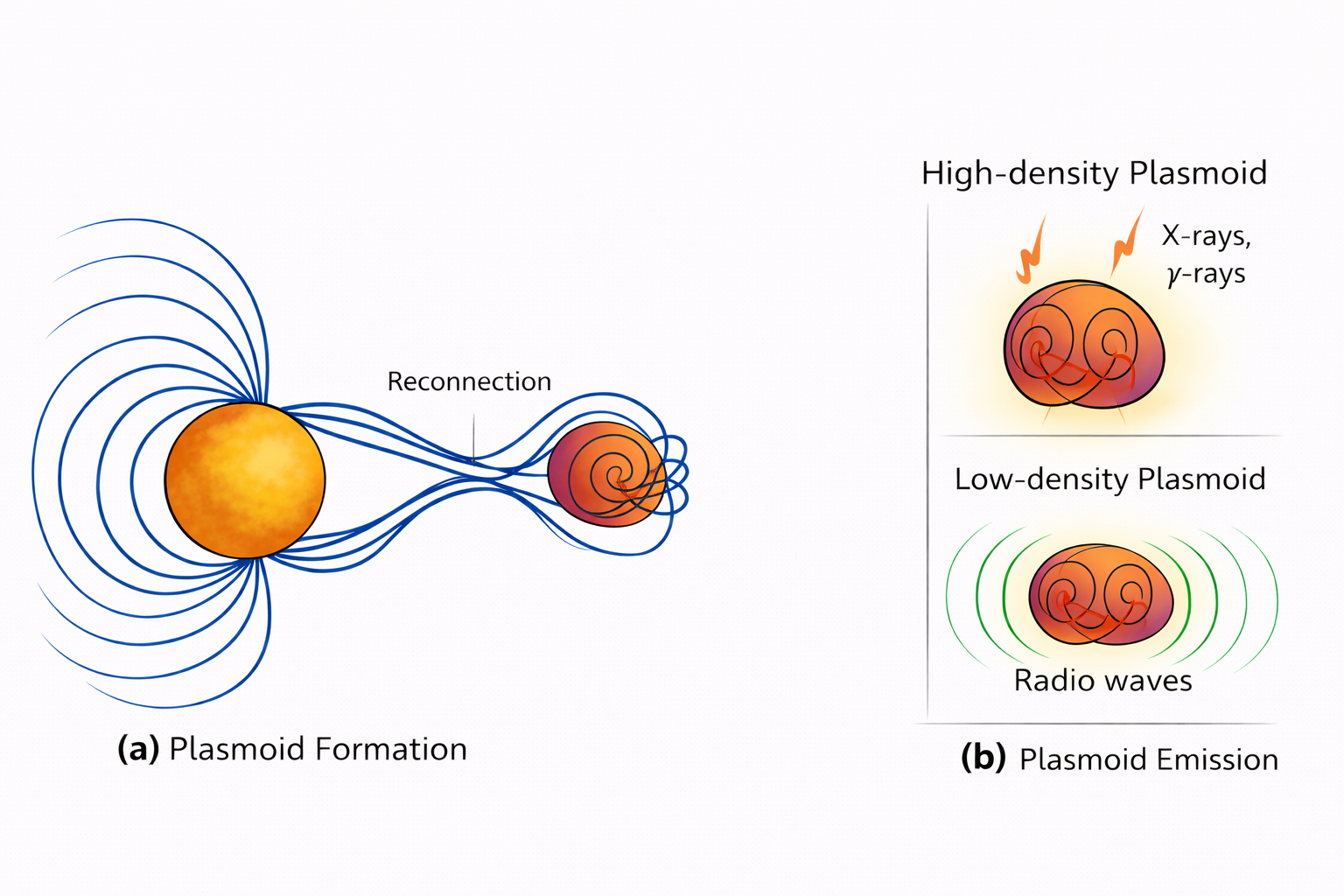}
    \caption{Schematic illustration of plasmoid formation and emission regimes associated with the different families of self-similar solutions.
(a) Magnetic reconnection in the magnetar magnetosphere leads to the formation of a magnetically confined plasmoid embedded within the closed-field region.
(b) Emission properties of plasmoids corresponding to different plasmoid regimes: P-type solutions (overdense) predominantly emit high-energy radiation (X-rays and $\gamma$-rays), whereas Z- and N-type solutions (underdense magnetically dominated configurations) correspond to low-density plasmoids capable of producing coherent radio emission.} 
    \label{fig:cartoon}
\end{figure}

The magnetic fields of magnetars are highly active and evolve due to MHD effects in the crust, such as the Hall drift \citep{Goldreich:1992,Pons:2009,Gourgouliatos:2016,Kojima:2021,Igoshev:2021}, and plastic flows \citep{Levin:2012,Lyutikov:2015,Lander:2016,Gourgouliatos:2021}, and to effects the core due to ambipolar diffusion and to the evolution of the superconducting magnetic field \citep{Graber:2015,Passamonti:2017b, Passamonti:2017a, Igoshev:2023,Skiathas:2024,Igoshev:2025}. These effects lead to the formation of multipolar magnetic structures in the lower magnetosphere and may release them to the magnetosphere. When an arcade forms in the magnetosphere, it might become unstable through high twist \citep{Parfrey:2013,Ntotsikas:2025} and can disconnect from the stellar magnetic field \citep{Lander:2024,Bransgrove:2025}. Such configurations evolve dynamically and give rise to a magnetar flare Fig.~ \ref{fig:cartoon}. Magnetar flares are rare explosive events that occur in magnetars. A flare can release immense amounts of energy that can reach $10^{46}$erg in the form of X-rays, shining within a fraction of a second. Moreover, flare ejecta expel masses reaching $10^{24}-10^{26}$ g \citep{Gaensler:2005,Granot:2006}, which subsequently power scaled-down versions of kilonovae through r-process radioactive decay \citep{Cehula:2024}.

In the model developed here, the relative content of the spherical configuration in mass, thermal pressure, magnetic and electric energy, temperature profile, and the maximum Lorentz factor attained are all related through the choices of the parameters $c_0$ and $c_1$. When we assume that a blob has a radius of $1$km ($10^{5}$ cm) and the maximum value of the magnetic flux corresponds to $10^{25}$ G cm$^{2}$, so that the magnetic field scales with $10^{15}$G, the magnetic and electric field energies $E_{mag}$ and $E_{el}$ range from $10^{44}$erg to $10^{46}$ erg. The pressure work $W$ ranges from $10^{43}$ erg to $10^{45}$ erg, and the temperature is at a level of $10^{11}$. We note that the temperature and energetics evolve with time as the plasmoid expands, with the decrease in electromagnetic and thermal energy of the sphere related to the work done against the external medium into which the sphere is expanding. The mass load $M$ ranges from $10^{24}$ g to $10^{25}$ g and does not evolve with time.

\subsubsection{Giant flare energetics}

We interpret magnetar giant flares as large-scale relativistic magnetic reconfigurations described by self-similar axisymmetric MHD solutions with finite mass loading. For all explored solution families (Z, P, and N), the energy budget at early times is dominated by the electromagnetic component, while the kinetic and thermal contributions depend on the mass loading and the pressure term, which are quantified through parameter $c_1$. This is a necessary condition for producing a bright, rapidly varying high-energy spike because an outflow dominated by kinetic or thermal energy would radiate inefficiently at X-ray and $\gamma$-ray energies and primarily power delayed shock-driven emission. The electromagnetic dominance of the early-time evolution is therefore consistent with the observation that the prompt hard X-ray or $\gamma$-ray spike contains the majority of the radiated energy in magnetar giant flares \citep[e.g.][]{Hurley:2005, Palmer:2005}.

An immediate observational implication is that giant flares can exhibit significant diversity in their late-time signatures despite comparable prompt luminosities. Solutions with higher mass loading (P-type) predict enhanced interaction between the ejecta and the surrounding medium, potentially giving rise to detectable radio afterglows or nebular disturbances, as observed following the 2004 flare of SGR~1806--20 \citep{Gaensler:2005}. In contrast, low-mass solutions (Z- and N-type) correspond to clean flares, in which little kinetic energy is available to power long-lived afterglow emission, consistent with the weaker radio signatures of SGR~1900+14 \citep{Frail:1999}.

\subsubsection{Timescales and light-curve morphology}

The self-similar expansion fixes the characteristic radial scale of the system to
\begin{equation}
r(t) = c v_0 ~ t ,
\end{equation}
with $v_0$ of order unity. Dissipation is expected to occur preferentially near the outer boundary of the expanding structure, where strong current sheets and pressure gradients develop. This naturally leads to sub-millisecond rise times for the initial hard X-ray or $\gamma$-ray spike, consistent with the observed rise times of Galactic giant flares \citep{Palmer:2005, Hurley:2005}.

The model predicts temporally asymmetric light curves, characterised by a very rapid rise phase followed by a longer decay. The rise time reflects localised dissipation at the expanding boundary, whereas the decay traces the subsequent global magnetic reconfiguration of the system. Since these timescales primarily depend on the expansion speed and not on the total energy release, similar rise times are expected across giant flares of different luminosities. While the presence of strong currents and discontinuities lead to dissipation and current-driven instabilities, a detailed study is required to confirm and follow the development of these effects in detail. 

\subsubsection{Angular structure and anisotropy}

The meridional structure of the solution leads to highly anisotropic profiles with magnetic field, pressure, and mass loading different on the equator and at the poles. Consequently, the emitted radiation is expected to be anisotropic, with the apparent energy of a flare depending on the observer’s viewing angle relative to the magnetic field symmetry axis.

This anisotropy provides a natural explanation for the wide dispersion in inferred isotropic-equivalent energies among observed giant flares \citep{Hurley:1999}, without requiring intrinsic energy releases spanning several orders of magnitude.

\subsubsection{Mass loading and flare diversity}

The explored parameter space spans ejecta masses ranging from effectively negligible values (Z-type solutions) to $\sim 10^{25}$--$10^{26}\,\mathrm{g}$. This range is consistent with estimates of the ejecta mass inferred from modelling the radio afterglow of SGR~1806--20 \citep{Gaensler:2005, Granot:2006}. Low-mass events, associated with Z- and N-type solutions, correspond to predominantly magnetospheric reconfigurations with minimum baryon contamination. These events are expected to produce sharp, luminous high-energy spikes and weak or undetectable afterglows.

In contrast, P-type solutions involve substantial mass loading, potentially including baryonic material originating from the stellar crust or inner magnetosphere \citep{Thompson:2001, Bransgrove:2025}. Such events are expected to exhibit stronger interaction with the ambient medium, leading to detectable radio afterglows or long-lived nebular emission.

\subsubsection{Spectral and polarisation signatures}

As the system expands self-similarly, the magnetic field strength decreases rapidly with radius, implying a fast spectral evolution of the emitted radiation. The model predicts an initially hard spectrum during the prompt spike, followed by rapid hard-to-soft evolution on millisecond timescales, in agreement with observed giant flare spectra \citep{Feroci:2001, Olive:2004}.

Because the magnetic field remains globally ordered during the early stages of the expansion, the prompt emission is expected to be highly linearly polarised, with a relatively stable polarisation angle during the rise phase. This is consistent with expectations for magnetically dominated emission regions \citep{Lyutikov:2006}. Significant depolarisation might only occur at later times, after magnetic reconnection and dissipation disrupted the large-scale field structure.

\subsection{Fast radio bursts}

Within the present framework, FRBs are interpreted as the coherent radio counterparts of magnetically dominated explosions with low baryon loading \cite{Lyubarsky:2021}. While the total energy release and large-scale dynamics can resemble those of magnetar giant flares, the conditions required for efficient coherent emission impose strong constraints on the ejecta mass and magnetisation. In particular, FRB production is favoured in solutions where the electromagnetic energy dominates the kinetic and thermal components throughout the early expansion \citep{Lyutikov:2017, Beloborodov:2017}.

In heavily mass-loaded explosions, such as P-type solutions, the inertia of the ejecta and the rapid formation of a strong compression shock suppress coherent emission by reducing the magnetisation and increasing the plasma frequency. These events are therefore expected to produce luminous high-energy flares, but weak or absent radio bursts. In contrast, Z- and N-type solutions correspond to minimally mass-loaded outflows in which the plasma remains highly magnetised and optically thin, allowing coherent radio emission to escape. This qualitative distinction is consistent with the observed diversity of magnetar activity, including radio-quiet giant flares and radio-loud but energetically modest bursts \citep{Kaspi:2017, Mereghetti:2020}.

The self-similar expansion naturally leads to the formation of strong current sheets and charge-starved regions near the expanding boundary of the magnetised structure. In these regions, large-amplitude electromagnetic disturbances can be generated on millisecond timescales, providing favourable conditions for coherent curvature radiation or synchrotron maser emission at relativistic shocks \citep{Lyubarsky:2014, Plotnikov:2019}. The characteristic duration of the resulting radio burst is set by the causal timescale $t \sim r/c$, which is consistent with the millisecond durations of observed FRBs.

An important prediction of the model is that FRBs need not be associated with the most energetic magnetar flares. Instead, they preferentially accompany low-mass explosions in which a large fraction of the released energy remains in electromagnetic form and is not degraded into kinetic energy at shocks \citep{Lyutikov:2022}. This provides a natural explanation for the observed association of FRBs with magnetars \citep{Popov:2013}, including the detection of a bright radio emission coincident with an X-ray flare from SGR~1935+2154 \citep{Cameron:2005,Bochenek:2020, Chime:2020}, and the absence of FRBs during the most extreme giant flares.

\section{Conclusions}
\label{sec:conclusions}

We have presented a class of self-similar axisymmetric MHD solutions describing relativistic magnetically dominated explosions with mass loading, motivated by energetic transient phenomena in magnetar magnetospheres. The solutions encompass a broad range of behaviours, characterised by different degrees of baryon loading and pressure support, while remaining analytically tractable.

A central result of this work is that the early-time evolution of these explosions is generically dominated by electromagnetic energy, with kinetic and thermal components becoming important only as a consequence of mass loading and interaction with the external medium. This regime naturally produces rapid high-energy emission on millisecond timescales, followed by a longer decay associated with magnetic energy depletion and eventual deceleration by a compression shock. The resulting light curves are strongly asymmetric, with rise times controlled primarily by the expansion speed and geometry and not by the total energy release.

Applied to magnetar giant flares, the model reproduces key observational properties, including sub-millisecond rise times, hard prompt spectra, low radiative efficiencies, and the presence or absence of afterglow emission, depending on the ejecta mass. Heavily mass-loaded solutions are associated with efficient energy transfer to shocks and long-lived radio nebulae, while lightly loaded solutions correspond to clean magnetospheric reconfigurations with weak or undetectable afterglows.

In this context, FRBs arise naturally from low-mass highly magnetised explosions in which coherent radio emission can escape before significant baryon loading or shock formation suppresses it. The model therefore predicts that FRBs need not accompany the most energetic magnetar flares, but instead preferentially occur in events where the electromagnetic energy remains dominant throughout the early expansion. This provides a unified explanation for the observed diversity of magnetar activity, including radio-quiet giant flares, radio-loud but energetically modest bursts, and rare events exhibiting both high-energy and radio emission. 

The present model is based on a semi-analytical approach, considering ideal MHD, axial symmetry, and self-similarity. These simplifications allow the consistent formation of models and the thorough exploration of the parameter space that illustrates the main qualitative characteristics. The qualitative behaviour of the solutions we presented is consistent with the numerical simulations of magnetically driven explosions in mass-loaded environments by \cite{Barkov:2022}. In particular, both approaches show that increasing the toroidal-to-poloidal magnetic field ratio leads to more compact configurations and a reduction in the maximum expansion velocity. Our semi-analytical solutions complement these numerical results by providing self-consistent equilibrium plasmoid structures that explicitly relate magnetic twist, plasma pressure, and inertia to the global dynamics of relativistic magnetic explosions. However, as these solutions are force-free while in our solutions, pressure and mass are central, a direct comparison is not yet possible. 

The key effects that are associated with observable properties of magnetar flares and FRBs require the relaxation of the assumed idealisations we used here. Namely, the expansion of a spherical magnetised plasmoid, as it propagates away from the neutron star, is not guaranteed to remain self-similar. Moreover, these spherical large-scale plasmoids can deviate from the spherical shape and axial symmetry. Finally, resistive instabilities and the possible formation of shocks due to the interaction of the mass with the external medium shape the radiative properties of the plasmoid. These deviations can be traced with a numerical simulation that will allow us to determine the limitations of the approach made here and their consequences for the observables.

Overall, the results presented here demonstrate that a single physically motivated class of magnetically dominated mass-loaded explosions can account for a wide range of observed transient phenomena associated with magnetars. By explicitly linking the dynamics to a small number of physical parameters, the model provides a coherent framework for interpreting current observations and for guiding future multi-wavelength and polarisation measurements of magnetar-driven transients.

\section*{Data availability}
\label{sec:Data Availability Statement}
The datasets generated during the current study are available from the author upon reasonable request.

\begin{acknowledgements}
The author acknowledges funding from grant FK 81641 "Theoretical and Computational Astrophysics", ELKE. This work was supported by computational time granted by the National Infrastructures for Research and Technology S.A. (GRNET S.A.) in the National HPC facility - ARIS - under project ID pr017008/simnstar2.
\end{acknowledgements}

\bibliographystyle{aa}
\bibliography{bibliography}

\end{document}